\documentclass[a4paper,12pt]{article}
\usepackage{amsmath}
\usepackage{amssymb}
\usepackage{braket}
\usepackage{cancel}
\usepackage{graphicx}
\usepackage{color}
\usepackage[utf8]{inputenc}
\usepackage{hyperref}
\usepackage{multirow}
\usepackage{slashed}
\usepackage{booktabs}
\usepackage[normalem]{ulem}
\usepackage{wasysym}
\usepackage{amssymb}
\usepackage{braket}
\usepackage{simplewick}
\usepackage{array}
\usepackage{graphicx}
\usepackage{cite}
\usepackage{float}
\usepackage{titlesec}
\usepackage{listings}
\usepackage{hyperref}
\usepackage{bbold}
\usepackage[english]{babel}
\usepackage{xcolor}
\usepackage{subcaption}
\usepackage{braket}
\usepackage{comment}
\allowdisplaybreaks

\usepackage{lscape}

\usepackage[english]{babel}

\usepackage{graphicx}

\usepackage{geometry}
\renewcommand{\arraystretch}{1.25}

\begin{document}

\begin{titlepage}

\begin{flushright}
{\small
P3H-22-090\\ 
SI-HEP-2022-21 \\
TTP22-056\\
Nikhef 2022-010 \\
%\today \\
%arXiv:20mm.nnnnn [hep-ph]
}
\end{flushright}

\vskip1cm
\begin{center}
{\Large \bf\boldmath 
New physics contributions to moments of inclusive \\ $b \to c$ semileptonic decays }
\end{center}

\vspace{0.5cm}
\begin{center}
{\sc Matteo Fael$^{a,b}$, Muslem Rahimi$^c$, 
and K. Keri Vos$^{d,e}$} \\[6mm]

{\it $^a$ Institut f{\"u}r Theoretische Teilchenphysik
Karlsruhe Institute of Technology (KIT), 76128 Karlsruhe, Germany\\[0.3cm]

{\it $^b$ Theoretical Physics Department, CERN, CH-1211 Geneva 23, Switzerland
}\\[0.3cm]

{\it $^c$ Center for Particle Physics Siegen (CPPS), \\
Theoretische Physik 1, Universit{\"a}t Siegen, \\
57068 Siegen, Germany}\\[0.3cm]

{\it $^d$Gravitational 
Waves and Fundamental Physics (GWFP),\\ 
Maastricht University, Duboisdomein 30,\\ 
NL-6229 GT Maastricht, the
Netherlands}\\[0.3cm]

{\it $^e$Nikhef, Science Park 105,\\ 
NL-1098 XG Amsterdam, the Netherlands}}
\end{center}

\vspace{0.6cm}
\begin{abstract}
\vskip0.2cm\noindent
Inclusive semileptonic $B\to X_c \ell \bar{\nu}_\ell$ decays, where $\ell = \mu,e$, are by now standard candles in the determination of the CKM element $|V_{cb}|$. These determinations rely on the heavy-quark expansion and use moments of decay spectra to extract the non-perturbative parameters directly from data under the standard model assumption. 

At the same time, new physics could influence the moments of the inclusive decay. 
In this paper, we compute power-corrections and next-to-leading order corrections in the strong coupling constant using the full basis of dimension-six new physics operators for the inclusive $B \to X_c \ell \bar{\nu}$ decay. We provide predictions for lepton energy, hadronic and leptonic invariant mass moments, and perform a phenomenological study to show the possible impact of new physics. Our results could be used to perform a global fit including new physics contributions. 
\end{abstract}

\end{titlepage}

\section{Introduction}
\label{sec::Introduction}
Semileptonic $b\to c$ decays provide important tests of the Standard Model (SM) of particle physics as they are mediated by a tree-level weak transition. As such, both the inclusive $B\to X_c \ell \bar{\nu}_\ell$ and exclusive $B\to D^{(*)}\ell \bar{\nu}_\ell$ decays, where $\ell = \mu,e$, are clean probes of the CKM element $|V_{cb}|$. For the exclusive decays, this requires information on the $B\to D^{(*)}$ form factors, while the inclusive decay relies fully on the heavy quark expansion (HQE) and the extraction of non-perturbative parameters from data. 
Thanks to a combined theoretical and experimental effort, the inclusive determination of $|V_{cb}|$ has reached an impressive  $1.2-1.5\%$ relative uncertainty \cite{Bordone:2021oof,Bernlochner:2022ucr}. 

Despite this progress, the puzzling tension between the exclusive and inclusive determination of $V_{cb}$ persists and has received quite some attention recently (see e.g.\ \cite{Grinstein:2017nlq,Bernlochner:2019ldg,Gambino:2019sif,Gambino:2020jvv,Bobeth:2021lya,FermilabLattice:2021cdg,Martinelli:2021myh}). 
At the same time the possible New Physics (NP) origin of this discrepancy has been investigated (see \cite{Colangelo:2016ymy,Jung:2018lfu, Crivellin:2014zpa,Crivellin:2009sd}). 
The search for such NP has been boosted by the recent finding of the $B$ anomalies, discrepancies between experimental data and theoretical SM predictions in both the neutral ($b\to s\ell\ell$) and charged ($b\to c \tau \bar{\nu}_\tau$) current decay of $B$ mesons. 

In this paper, we consider the effect of possible new physics interactions 
on moments of the inclusive $B\to X_c \ell \bar\nu_\ell$ decay, for light leptons.
The effect of NP on the moments of the $b\to c$ spectrum have so far only been studied in \cite{Mannel:2017jfk, Dassinger:2008as}, where a subset of possible NP operators was included. 
NP contributions to the total inclusive rate were included in the analysis of Ref.~\cite{Jung:2018lfu}, while new tensor interactions were discussed in \cite{Colangelo:2016ymy}. 

Using the framework of the HQE, we consider the $B\to X_c\ell\bar \nu_\ell$ spectra including the full set of NP dimension-six operators appearing in the weak effective theory (WET) below the electroweak (EW) scale.
We provide predictions for lepton energy ($E_\ell$), hadronic ($M_X^2$) and leptonic ($q^2$) invariant mass moments. Moreover we study also NP effects in forward-backward asymmetries which were proposed in~\cite{Turczyk:2016kjf} and recently reconsidered in~\cite{Herren:2022spb}.
When considering the most general effective Hamiltonian for $b\to c \ell \bar \nu_\ell$ transition with dimension-six operators, we have three expansion parameters in the HQE: the inverse of the EW scale $G_F= 1/(\sqrt{2}v^2)$, $1/m_b$ and $\alpha_s(m_b)$.
In order to properly catch the leading effects in the various moments, 
we compute the following kind of contributions:
\begin{itemize}
    \item NP contributions at tree level in the free-quark approximation. 
    These terms scale like $G_F^2 \times \alpha_s^0 \times (\frac{1}{m_b})^0$
    in the prediction for the differential rate. 
    Note that the interference between SM and NP operators vanishes
    for scalar and tensor currents when the leptons are considered massless.
    \item Power suppressed contributions up to order $1/m_b^3$ also for the 
    NP operator contributions. 
    These corrections scale like
    $G_F^2 \times \alpha_s^0 \times (\frac{1}{m_b})^{2,3}$.
    Since the prediction for $q^2$ and $M_X^2$ central moments receive large 
    contributions from power corrections, it is important to consider also 
    the power suppressed terms for the NP effects.
    \item Perturbative QCD NLO corrections to the NP effective interactions in the
    free quark approximation, which 
    scale like $G_F^2 \times \alpha_s^1 \times (\frac{1}{m_b})^{0}$.
    For the second and third central moments of $M_X^2$, the $\alpha_s$ corrections
    are much larger than the partonic contribution. Because the partonic invariant mass differs only from $m_c$ starting at $\mathcal{O}(\alpha_s)$, the NLO corrections are effectively the LO contribution.
\end{itemize}
In the end, our results could be included in a fit to the experimental data to constrain possible NP contributions. We plan to implement this in the EOS software \cite{EOSAuthors:2021xpv}. In the mean time, to show the impact of such an analysis, we illustrate the effect of different NP scenarios with some phenomenological studies. Finally, we present a toy fit to show the effect on the $V_{cb}$ extraction, as the HQE parameters could mimic the effect of NP. 

This work is organised as follows. In Section~\ref{sec::Theory} we introduce the set
of dimension-six operators which can contribute to the inclusive semileptonic $B$ decay and discuss the derivation of the NLO corrections for the NP operators. In Sec.~\ref{sec:NPmoments} we present the results for the NP contributions 
to moments, illustrate their effects using three benchmark scenarios and study their impact on the extraction of the HQE parameters in global fits via a toy fit. In Sec.~\ref{sec:AFB} we discuss the effects of NP in the forward-backward asymmetries. We conclude in Sec.~\ref{sec::Conclusion}. In Appendix~\ref{app:nptot}, we give the contribution to the total rate, while in Appendix~\ref{sec:appxi} we give our results for the different contributions to the moments. 

\section{\boldmath Effective NP contributions to $b\to c \ell \bar{\nu}_\ell$}
\label{sec::Theory}
We consider NP effects in $b\to c\ell\bar{\nu}_\ell$ decays arising from 
\begin{align}
    \mathcal{H}_{\text{eff}} &= \frac{4 G_F V_{cb}}{\sqrt{2}} 
    \left [ \left(1+ C_{V_L} \right) O_{V_L} 
    + \sum_{i = V_R, S_L, S_R, T} C_{i} \, O_{i}  \right],
    \label{eq::Hamiltonian-NP}
\end{align}
where the effective dimension-six operators are 
\begin{align}
    O_{V_{L(R)}} &= \left( \bar{c} \gamma_\mu P_{L(R)} b \right) \left(\bar{\ell} \gamma^\mu P_{L} \nu_\ell \right) \, , \\
    O_{S _{L(R)}} &= \left(\bar{c} P_{L(R)} b \right) \left(\bar{\ell} P_{L} \nu_\ell \right) \, \\
    O_{T} &= \left(\bar{c} \, \sigma_{\mu \nu} P_{L} b \right) \left(\bar{\ell} \, \sigma^{\mu \nu} P_{L} \nu_\ell \right) \, .
\end{align}
with $P_{L(R)} = 1/2 \, (1 \mp \gamma_5)$ and $\sigma^{\mu \nu} = \frac{i}{2} [\gamma^\mu, \gamma^\nu]$. In the SM only $O_{V_{L}}$ contributes. We have written out this contribution explicitly, such that all Wilson coefficients $C_i$ are zero in the SM. We do not consider interactions with right handed neutrinos (see e.g. \cite{Mandal:2020htr} for a discussion of these effects on exclusive $B\to D^{(*)} \ell \bar{\nu}_\ell$ decays). 

Note that if one would consider NP effects in the SMEFT framework~\cite{Grzadkowski:2010es}, 
there would be an additional expansion in powers of $1/\Lambda$, where
$\Lambda$ corresponds to the NP scale above the EW scale. 
The tree-level matching of SMEFT operators onto the effective Hamiltonian
can be obtained from~\cite{Aebischer:2015fzz}.
In the WET the expansion parameter is $1/v$, therefore from the SMEFT point of view
the Wilson coefficients in Eq.~\eqref{eq::Hamiltonian-NP} would be further 
suppressed by the small ratio $(v/\Lambda)^2$.

To study the effects of the NP operators on moments of the spectrum, we calculate the triple differential decay rate in terms of the lepton (neutrino) energy $E_{\ell (\nu)}$ and the dilepton invariant mass $q^2 = (p_\ell + p_\nu)^2$. We write
\begin{align}
\label{eq:diff}
     \frac{\text{d}\Gamma_{\text{SM+NP}}}{\text{d}E_{\ell} \text{d}q^2 \text{d}E_{\nu}} &= \frac{G_{F}^2 |V_{cb}|^2} {16 \pi^3}  \tilde{W} \otimes \tilde{L}\ , 
\end{align}
where  
\begin{align}
     \tilde{W} \otimes \tilde{L} &\equiv  |1 + C_{V_{L}}|^2 \left( W_{\mu \nu} L^{\mu \nu} \right)_{V_{L},V_{L}} + |C_{V_{R}}|^2 \left( W_{\mu \nu} L^{\mu \nu} \right)_{V_{R},V_{R}} + |C_{S_{L}}|^2 \left( W L \right)_{S_{L},S_{L}} \nonumber \\
    & + |C_{S_{R}}|^2 \left( W L \right)_{S_{R}, S_{R}} 
    + |C_{T}|^2 \left(W_{\mu \nu \rho \sigma} L^{\mu \nu \rho \sigma} \right)_{T, T} +  \text{Re} ( (1+ C_{V_{L}} ) C_{V_{R}}^* ) \left (W_{\mu \nu} L^{\mu \nu} \right)_{V_{L}, V_{R}} \nonumber \\
    & + \text{Re}( C_{S_{L}} C_{S_{R}}^{*} ) \left (W L \right)_{S_{L},S_{R}} + \text{Re}(C_{S_L} C_T^*) (W_{\mu \nu} L^{\mu \nu})_{{S_L},T} \nonumber \\
    & + \text{Re}(C_{S_R} C_T^*) (W_{\mu \nu} L^{\mu \nu} )_{{S_R},T} \ .
    \label{eq:WtimesL}
\end{align}
We split the contributions into the lepton ($L$) and hadronic ($W$) tensors. 
We define
\begin{align}
    L &= 
    \sum_{\rm{lepton \; spin}} 
    \bra{0}J_{L}^{\dagger} \ket{\ell\bar{\nu}_\ell}
    \bra{\ell \bar{\nu}_\ell} J_{L'} \ket{0} \ , 
    \label{eq::LeptonTensor}
\end{align}
where we suppressed the Lorenz indices in the leptonic tensor.
The indices $L$ and $L'$ can take the values $S_{L,R}, V_{L,R}$ and $T$ with 
\begin{align}
    J_{S_{L,R}} &= 
    (\bar \ell P_L \nu_\ell), &
    J_{V_{L,R}}^\mu &= 
   ( \bar \ell \gamma^\mu P_L \nu_\ell),&
    J_{T}^{\mu\alpha} &= 
    (\bar \ell \sigma^{\mu\alpha} P_L \nu_\ell).
\end{align}
We define the hadronic tensor in the following way:
\begin{align}
    W &= 
    \sum_{X_c} \frac{1}{2 m_{B}} (2\pi)^3 
    \bra{\bar{B}}J_{H}^{\dagger} \ket{X_{c}} \bra{X_{c}} J_{H'} \ket{\bar{B}} \delta^{(4)}(p_{B} -q - p_{X_{c}}) \ , 
    \label{eq::HadronTensor}
\end{align}
where $p_{X_c}$ is the total momentum of the $X_c$ state 
and also in this case we suppressed the Lorenz indices. 
In the presence of NP interactions, the index $H$ and $H'$ can take the values $S_{L,R}, V_{L,R}$ and $T$ where
\begin{align}
    J_{S_{L(R)}} &= (\bar{c} P_{L(R)} b) \ , &
    J^{\mu}_{V_{L(R)}} &= (\bar{c} \gamma^\mu P_{L(R)} b) \ , &
    J_T^{\mu \alpha} &= (\bar{c} \sigma^{\mu \alpha} P_L b) \ .
    \label{eqn:hadroniccurrents}
\end{align}
In Eq.~\eqref{eq:WtimesL} we neglected combinations of the form $\left (W_{\mu } L^{\mu} \right)_{V_{L(R)}, S_{L(R)}}$ and $\left (W_{\mu\rho\sigma } L^{\mu\rho\sigma} \right)_{V_{L(R)}, T}$  since they do not contribute in the limit $m_\ell \to 0$ considered in this work. The hadronic tensors $W$ can now be calculated using the heavy quark expansion (HQE) (see e.g.~\cite{Manohar:2000dt}), expressing them in pertubatively calculable coefficients and hadronic matrix elements scaling with inverse powers of $m_b$. The number of matrix elements proliferates at each higher order in $1/m_b$ (see \cite{Dassinger:2006md,Mannel:2018mqv,Fael:2018vsp}). Here we only consider terms up to $1/m_b^3$ defined as: (see e.g.~\cite{Mannel:2010wj}) 
\begin{eqnarray}
2 m_B \, (\mu_\pi^2)^\perp & \equiv & 
- \langle B |\bar b_v (iD_\rho) (iD_\sigma) b_v| B \rangle \Pi^{\rho\sigma} \ ,
\notag
\\
2 m_B \, (\mu_G^2)^\perp & \equiv & \frac{1}{2}
\langle B |\bar b_v \left[i D_\rho, i D_\lambda\right] (- i \sigma_{\alpha \beta}) b_v| B \rangle  \Pi^{\alpha \rho}  \Pi^{\beta\lambda},
\notag
\\
2 m_B \, (\rho_D^3)^\perp & \equiv & \frac{1}{2}
\langle B |\bar b_v \left[i D_\rho, \left[iD_\sigma, iD_\lambda\right]\right] b_v| B \rangle \Pi^{\rho\lambda}  v^\sigma,
\notag
\\
2 m_B \, (\rho_{LS}^3)^\perp & \equiv & \frac{1}{2}
\langle B |\bar b_v\left\{i D_\rho, \left[iD_\sigma, iD_\lambda\right]\right\}(-i\sigma_{\alpha\beta}) b_v| B \rangle \Pi^{\alpha\rho}\Pi^{\beta\lambda}  v^\sigma\, ,
\end{eqnarray}
where $v^\mu = p_B^\mu/m_B$ is the velocity of the $B$ meson and
\begin{equation}
    \Pi_{\mu\nu} = g_{\mu\nu}-v_{\mu}v_\nu \ . 
\end{equation}
In the following, we drop the ``perp'' superscript for simplicity. Alternative, the HQE parameters can be defined with the full covariant derivative, related to the spatial component via $iD^\mu= v^\mu (iv \cdot D) + D_\perp^\mu$. 
These definitions were used in Refs.~\cite{Mannel:2018mqv, Fael:2018vsp, Bernlochner:2022ucr} as, in the reparametrization invariant (RPI) basis, it is beneficial to use the full derivative  (see discussion in Appendix~A of \cite{Fael:2018vsp} for the relation between these two bases). 
In principle, the $1/m_b^4$ terms can be included as recently done for the $q^2$ moment analysis \cite{Bernlochner:2022ucr}. The two $1/m_b^4$ parameter extracted were found to be consistent with zero. These higher-order corrections were also studied in \cite{Gambino:2016jkc} using the lowest-lying state approximation. Therefore, for this study of NP effects, we only consider terms up to $1/m_b^3$. 

\subsection{Next-to-leading order corrections}
\label{sec::NLO}
Besides these power-corrections, we also compute the NLO corrections to the triple differential rate for the full NP operator basis in (\ref{eq::Hamiltonian-NP}). For scalar NP interactions, the NLO corrections to the $q^2$ spectrum are already given in \cite{Celis:2016azn}, using results from \cite{Czarnecki:1997sz}. 
The NLO corrections for the SM are well known for both the massive and massless leptons in the semileptonic decay $b \to c \ell \bar{\nu}_\ell$ \cite{Jezabek:1988iv,Jezabek:1996db,DeFazio:1999ptt, Trott:2004xc,Aquila:2005hq,Gambino:2005tp, Bauer:2002sh}. 

\begin{figure}[t]
	   \centering
        \subfloat[]{\includegraphics[width=0.30 \textwidth]{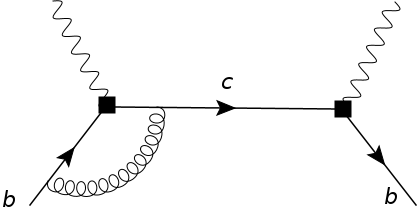}}\hspace{1cm}
        \subfloat[]{\includegraphics[width=0.30\textwidth]{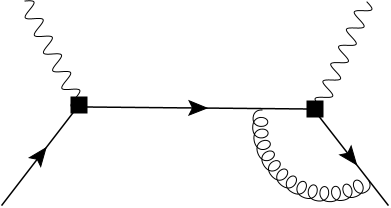}}\\
        \vspace{0.5cm}
        \subfloat[]{\includegraphics[width=0.30\textwidth]{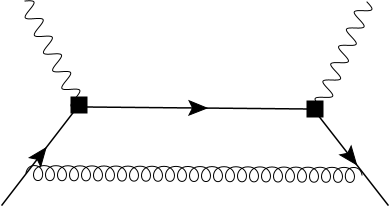}}\hspace{1cm}
        \subfloat[]{\includegraphics[width=0.30\textwidth]{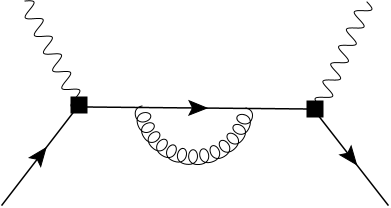}}
      \caption{One-loop forward scattering diagrams which contribute to the $b \to X_c \ell \bar \nu_\ell$ differential rate at NLO.
      The black boxes represents one of the currents $J_{H}$ defined in ~\eqref{eqn:hadroniccurrents}. 
      Solid lines represent the quarks, curly lines the gluons and wavy lines 
      the color-singlet external current mediating the weak decay.}
        \label{fig::oneloop-forward}
\end{figure} 
We compute the $\mathcal{O}(\alpha_s)$ for the structure functions of the hadronic tensor $W$ for the different currents which enter the fully differential decay width. We note that it turns out to be more convenient to express the triple differential rate with respect to $u \equiv p_{X_c}^2- m_c^2$ instead of $E_\nu$ as in \cite{Aquila:2005hq}. We then extract the predictions for the various moments and forward-backward asymmetries with arbitrary cuts via numerical integration of the differential rate
over the allowed phase space, following the approach described in~\cite{Alberti:2016fba}.

In general we can express the structure functions as:
\begin{align}
    W_{HH'}(q^2, (v\cdot q)) &= 
    W_{HH'}^{(0)}(q^2, (v\cdot q)) 
    + \frac{\alpha_s(\mu)}{\pi} 
    \Big[
    W^{(1)}_{HH', \mathrm{virt}}(q^2, (v\cdot q)) 
    +W^{(1)}_{HH', \mathrm{real}}(q^2, (v\cdot q)) 
    \Big] \, ,
\end{align}
where ``virt'' and ``real'' stand for virtual and real contributions, respectively. 
The indices $HH'$ run over all possible pairs of NP interactions, e.g.\ $V_L V_R, 
S_L S_R$, etc.

For the ultraviolet and infrared divergences, we use dimensional regularization 
and define $\epsilon = (4-d)/2$, where $d$ is the space-time dimensions.
For the calculation we use the Mathematica package \texttt{FeynCalc}~\cite{Shtabovenko:2020gxv}.
The ultraviolet divergences in the one-loop virtual diagrams are removed by using on-shell quark mass and wave function renormalization. 
Furthermore, there are additional ultraviolet divergences for the scalar and tensor currents.  We therefore apply a renormalization of these currents
according to their one-loop anomalous dimension (see e.g.~\cite{Aebischer:2017gaw}). 
For the computation of real emission we employed the inverse unitarity approach~\cite{Anastasiou:2002yz}. 
This method allows us to rewrite the real emission diagram integrated over the gluon phase-space as a multi-loop integral with cut propagators.
We can then apply the usual IBP reduction to reduce the real emission contribution to phase-space master integrals which are then calculated explicitly. 
In the process of the reduction to master integrals we take into account the cut in the gluon and charm intermediate state. 
For the real emission we encounter the following integral family:
\begin{align}
    I(a,b,c) &= \left( 4\pi \, e^{-\gamma_E} \right)^{-\epsilon} 
    \mathrm{Disc}
    \int \frac{\text{d}^d k}{(2\pi)^d} \frac{1}{[k^2]^a [(p_b -k)^2 -m_b^2]^b [(p_b-q-k)^2-m_c^2]^c} 
\end{align}
By applying the Cutkosky's rules for the gluon and charm intermediate state:
\begin{align}
    \frac{1}{k^2} &\to (-2\pi i ) \, \delta(k^2) \, , \notag \\
    \frac{1}{(p_b -q -k)^2 -m_c^2} &\to (-2\pi i ) \, \delta((p_b-q-k)^2 -m_c^2) \, ,
\end{align}
we obtain the following master integrals:
\begin{align}
    I(1,0,1) &= 
    \left( 4\pi \, e^{-\gamma_E} \right)^{-\epsilon} 
    \int \frac{\text{d}^d k}{(2\pi)^d} \,
    (-2\pi i)^2 \delta(k^2) \delta((p_b-q-k)^2-m_c^2) \Theta(k_0) \nonumber \\
     &= 
     -\frac{\hat{u}}{4\pi \hat s} 
     \left(\frac{\hat{u}}{\sqrt{\hat s}}\right)^{-2 \epsilon} 
     \left(\frac{1}{2} +\epsilon + \mathcal{O}(\epsilon^2)\right) \, , \\
     I(1,1,1) &= \left( 4\pi \, e^{-\gamma_E} \right)^{-\epsilon} 
     \int \frac{\text{d}^d k}{(2\pi)^d} \, \frac{1}{(p_b+k)^2-m_b^2} (-2\pi i)^2 \delta(k^2) \delta((p_b-q-k)^2-m_c^2) \Theta(k_0) \nonumber \\
    &=  
   \left(\frac{\hat{u}}{\sqrt{\hat s}}\right)^{-2 \epsilon} 
   \left\{
  \frac{1}{8 \pi 
   \sqrt{\lambda }}
   \log \left(
   \frac{1-\hat q^2 +\hat s + \sqrt{\lambda }}
   {1-\hat q^2 + \hat s - \sqrt{\lambda }}
   \right)   
   \right.
   \notag \\ &
   \left.
     +\frac{\epsilon}{4\pi \sqrt{\lambda}}
     \left[
          \text{Li}_2
          \left(
          \frac{2 \sqrt{\lambda}}
          {1-\hat q^2+\hat s +\sqrt{\lambda}}
          \right) 
   +\frac{1}{4}
   \log^2
   \left(
   \frac{1-\hat q^2 + \hat s + \sqrt{\lambda }}
   {1-\hat q^2 + \hat s - \sqrt{\lambda }}
   \right) 
   \right] 
   +\mathcal{O}(\epsilon^2)
   \right\} \, ,
\end{align}
where $\hat s = \rho+\hat u$, $\lambda = \lambda(1,\hat{q}^2, \hat s) $ and 
$\lambda(x,y,z) = x^2+y^2+z^2-2xy -2xz -2yz$ is the K\"allen function.
The singularities of the real emissions are located at $\hat{u} = 0$ with:
\begin{align}
    \hat{u} &=  (1-\hat{q})^2 -\rho \,, \hspace{1cm} 0 \leq \hat{u} \leq \hat{u}_{\text{max}}=(1-\sqrt{\hat{q}^2})^2 -\rho  \, .
\end{align}
We have to extract the singular behavior of the master integrals around $\hat u = 0$ before expanding in $\epsilon$.
The infrared divergences are extracted explicitly by using the plus distribution:
\begin{align}
    \hat{u}^{-1+a \epsilon} &= \frac{1}{a \epsilon} \delta(\hat{u})\, \hat{u}_{\text{max}} + \left[ \frac{1}{\hat{u}} \right]_{+} + \mathcal{O}(\epsilon) \, .
\end{align}
The integration of the plus distribution over a test function is defined as:
\begin{align}
    \int_0^{\hat u_\mathrm{max}} f(\hat{u}) \left[ \frac{1}{\hat{u}}\right]_+ \text{d}\hat{u} &= \int_0^{\hat u_\mathrm{max}} \frac{ f(\hat{u}) -f(0)}{\hat{u}} \text{d}\hat{u} \, .
\end{align}
In the sum between real and virtual corrections all the infrared divergences cancel.
For the $\gamma_5$ definition in dimensional regularization we use the Larin prescription \cite{Larin:1993tq}, i.e.\
\begin{align}
    \gamma_5 &= \frac{i}{12} \epsilon_{\mu_1 \mu_2 \mu_3 \mu_4} \gamma^{\mu_1} \gamma^{\mu_2} \gamma^{\mu_3} \gamma^{\mu_4},
\end{align}
which requires an additional finite renormalization constant in
order to restore the correct Ward identity.

Note that, our method to compute the one-loop diagrams differs from \cite{Aquila:2005hq} where they regularize IR divergences via a finite gluon mass. 
Ref.~\cite{Aquila:2005hq} presented also the corrections of $\mathcal{O}(\alpha_s^n \beta_0^{n-1})$ (the so-called large-$\beta_0$ limit). 
This can be also done in our approach, however, we do not include them in this analysis. 
To summarize, in this work we consider leading order, power-corrections up to $\mathcal{O}(1/m_b^3)$ and next-to-leading order corrections. Schematically:
\begin{align}
    \frac{\text{d}\Gamma_{\text{SM+NP}}}{\text{d}E_\ell \text{d}q^2 \text{d}E_\nu} &=  \frac{\text{d}\Gamma^{\text{LO}}_{\text{SM+NP}}}{\text{d}E_\ell \text{d}q^2 \text{d}E_\nu} +  \frac{\text{d}\Gamma^{\text{Pow}}_{\text{SM+NP}}}{\text{d}E_\ell \text{d}q^2 \text{d}E_\nu} + \left(\frac{\alpha_{s}}{\pi} \right) \frac{\text{d}\Gamma^{\text{NLO}}_{\text{SM+NP}}}{\text{d}E_\ell \text{d}q^2 \text{d}E_\nu} \, .
    \label{eq:WLcontraction}
\end{align}

\subsection{Moments of the spectrum}
In the following, we consider the lepton energy moments, dilepton invariant mass ($q^2$) moments and hadronic invariant mass moments of the $b\to c$ spectrum. The first two can be easily obtained from the triple differential rate defined as in \eqref{eq:diff}. The hadronic invariant mass is related to these variables via
\begin{align}
M_{X}^{2} &\equiv (p_B - q)^2 = (m_B^2 + q^2 -2 m_B (v\cdot q)) \ .
\label{eq::Mx1}
\end{align}
The normalized moments for observable $\mathcal{M}$ are then defined
\begin{align}
    \label{eq:moments}
    \braket{\mathcal{M}^{n}}_{E_\ell > E_\ell^{\text{cut}}} &= \frac{\int_{E_\ell > E_\ell^{\text{cut}}} \text{d}\mathcal{M} \, \mathcal{M}^n \frac{\text{d} \Gamma}{\text{d} \mathcal{M}} } {\int_{E_\ell > E_\ell^{\text{cut}}} \text{d}\mathcal{M} \, \frac{\text{d} \Gamma}{\text{d} \mathcal{M}} },
\end{align}
where $E_\ell^{\text{cut}}$ is the energy cut of the lepton $\ell = (e, \, \mu)$ and $n$ denotes the $n$-th order of moment. Similarly, for $q^2$ moments, we consider moments with minimum cut $q^2_\mathrm{cut}$ on the value of $q^2$. 
As is customary, we also calculate central moments defined as
\begin{align}
\label{eq:central-moments}
\braket{ (\mathcal{M} -  \braket{\mathcal{M}})^{n}} &= \sum_{i=0}^{n} \, \left(\begin{array}{c} n \\ i  \end{array}\right)\braket{(\mathcal{M})^i} (- \braket{\mathcal{M}})^{n-i} % \text{where} \hspace{0.5cm} a = \braket{\mathcal{O}} 
\, .
\end{align}
The moments can be obtained using Eq.~\eqref{eq:moments} and by integrating the triple differential rate over the allowed phase space.

\section{\boldmath New physics in moments of $B\to X_c \ell \bar{\nu}_\ell$}
\label{sec:NPmoments}

The moments can now be obtained from the triple differential rate in \eqref{eq:diff}. We write 
\begin{align}
 \braket{\mathcal{M}} &= \xi_{\text{SM}} +  |C_{V_{R}}|^2 \, \xi^{\braket{V_R,V_R}}_{\text{NP}} +  |C_{S_{L}}|^2 \, \xi^{\braket{S_L,S_L}}_{\text{NP}} +|C_{S_{R}}|^2 \, \xi^{\braket{S_R,S_R}}_{\text{NP}} + |C_{T}|^2 \, \xi^{\braket{T,T}}_{\text{NP}}  \nonumber \\ 
 &  + \text{Re}(( C_{V_{L}} -1) C_{V_R}^*) \, \xi^{\braket{V_L,V_R}}_{\text{NP}}  + \text{Re}(C_{S_{L}} C_{S_R}^*) \, \xi^{\braket{S_L , S_R}}_{\text{NP}} + \text{Re}(C_{S_{L}} C_{T}^*) \, \xi^{\braket{S_L , T}}_{\text{NP}} \nonumber \\
 & + \text{Re}(C_{S_{R}} C_{T}^*) \, \xi^{\braket{S_R , T}}_{\text{NP}} \ ,
 \label{eq::moments}
\end{align}
where we assume that the NP Wilson coefficients are smaller than one so that
we can expand the ratios in Eq.~\eqref{eq:moments} up to quadratic NP couplings. Earlier calculations of the triple differential rate at tree-level including some NP contributions can also be found in \cite{Goldberger:1999yh}.
The contribution $C_{V_L} \xi^{\braket{V_L}}_{\text{NP}}$ drops out for normalized moments and in the branching ratio it is equivalent to a rescaling of $V_{cb}$. 
The coefficients denoted by $\xi$ depend on the bottom and charm quark masses, the HQE parameters and the lepton energy cut or the $q^2$ cut. 
For $\xi_{\rm SM}$, we agree with the numerical results at $\mathcal{O}(\alpha_s)$ given in \cite{Gambino:2011cq} for the electron energy and $M_X$ moments in the kinetic scheme. 
The NP coefficients with NLO corrections are lengthy and require numerical integration depending on the lepton energy (or $q^2$) cut. Therefore, we do not report explicitly our results. They can be obtained in Mathematica format from the web repository
\texttt{https://gitlab.com/vcb-inclusive/npinb2xclv}.
The content of the repository is presented in Appendix~\ref{sec:git}.
However, to illustrate the effect of the NP contributions, we report our predictions for the various central moments for benchmark values of the cuts. We consider $E^{\text{cut}}_{\ell} = 1$ GeV in case of the lepton energy and hadronic invariant mass moments. For the $q^2$ moments, we present results for $q^2_{\rm cut}=4$ GeV$^2$. In the next section, we also illustrate the lepton energy or $q^2$ cut dependence for specific NP scenarios. 

In Appendix~\ref{sec:appxi} we report our predictions for the different moments. 
We work in the kinetic scheme~\cite{Bigi:1996si,Czarnecki:1997sz,Fael:2020iea,Fael:2020njb}. 
We fix the value of the scale $\mu$ in $m_b^{\rm kin}(\mu)$ at 1~GeV. 
For the charm quark mass we use the $\overline{\text{MS}}$ scheme and fix $\overline{m}_c$(2 GeV). 
For the strong coupling constant we use $\alpha_{s}(m_{b}^\mathrm{kin})= 0.2184$ \cite{Herren:2017osy}. 
In addition, we use the input values in Table \ref{table:input}. 
These are obtained from a global fit to lepton energy and hadronic invariant mass moments of the $B\to X_c \ell \bar{\nu}_\ell$ spectra in~\cite{Bordone:2021oof} (which updates the fit of \cite{Gambino:2016jkc}). Interestingly, the value of $\rho_D^3$ in Table~\ref{table:input} differs from the determination of $\rho_D^3 = (0.03\pm0.02)\,  \, \text{GeV}^3$ found in \cite{Bernlochner:2022ucr}. 
The latter uses $q^2$ moments, which depend on a reduced RPI basis of HQE elements. Specifically, $\rho_{LS}^3$ does not enter into the prediction of RPI quantities and the dependence on $\mu_\pi^2$ is very much reduced for normalized $q^2$ moments.
The difference between the values for $\rho_D^3$ obtained from these two data sets requires further study, preferably via a combined fit to all available data. These studies are in progress. 
On the contrary, the lepton and hadronic mass moments depends on $\rho_{LS}^3$ and $\mu_\pi^2$, so we cannot use the HQE parameter values from \cite{Bernlochner:2022ucr} for these moments. However, for the $q^2$ moments both determinations of HQE parameters can be used. We comment on this in the next section.

\begin{table}[t]
\begin{center}
\begin{tabular}{ c | c }
\hline\hline
$m^{\text{kin}}_{b}$ & (4.573 $\pm$ 0.012) GeV 
\\
\hline
$\overline{m}_{c}$(2 GeV) & (1.092 $\pm$ 0.008) GeV  
\\ 
\hline
$\left(\mu_{\pi}^{2}(\mu)\right)_{\text{kin}}$ & (0.477 $\pm$ 0.056) GeV$^{2}$
\\
\hline
$\left(\mu_{G}^{2}(\mu)\right)_{\text{kin}}$ & (0.306 $\pm$ 0.050) GeV$^{2}$ 
\\
\hline
$\left(\rho_{D}^{3}(\mu)\right)_{\text{kin}}$ & (0.185 $\pm$ 0.031) GeV$^{3}$
\\
\hline
$\left(\rho_{LS}^{3}(\mu)\right)_{\text{kin}}$ & (-0.130 $\pm$ 0.092) GeV$^{3}$
\\
\hline\hline
\end{tabular}
\end{center}
\caption{Numerical inputs from \cite{Bordone:2021oof}. The HQE parameters and the $b$-quark mass are given in the kinetic scheme at $\mu = 1$ GeV.}
\label{table:input}
\end{table}

Our results in Appendix~\ref{sec:appxi} show the impact of different NP contributions. As stated already in the introduction, especially for the $M_X$ and $q^2$ moments, the inclusion of $1/m_b$ power corrections is crucial, while in addition for the former also $\alpha_s$ numerically plays an important role. 
In principle, the coefficients have an uncertainty stemming from the input parameters. However, here we refrain from giving those. We include them in the next section when discussing different NP scenarios.

\begin{figure}[t]
	\centering
	{\includegraphics[width=0.5\textwidth]{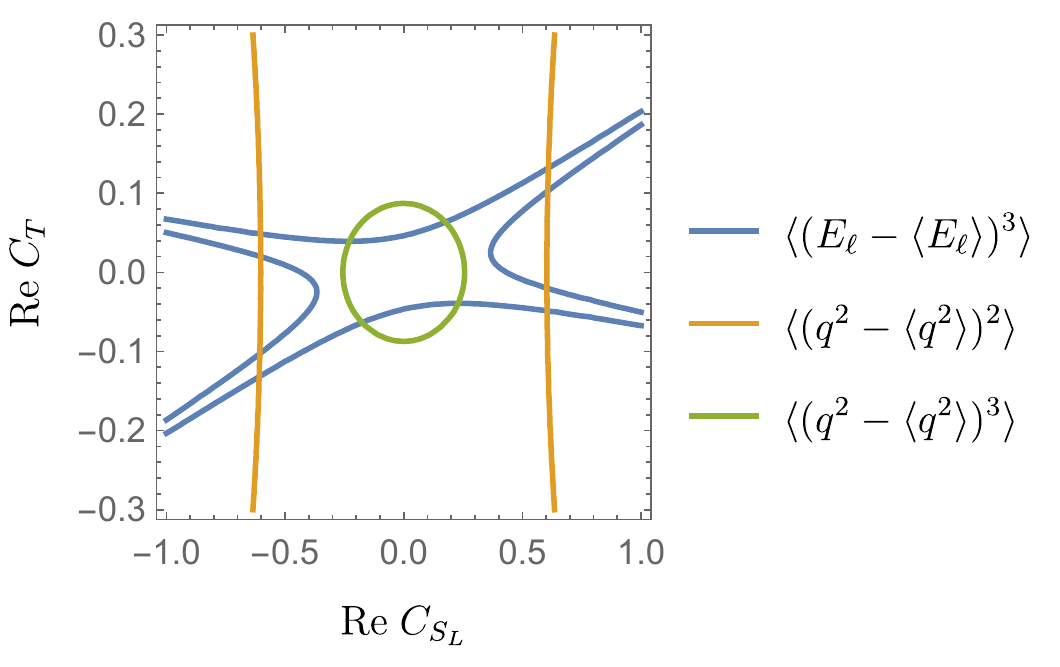}}
  \caption{Illustration of complementarity of constraints on $C_{S_L}$ and $C_T$ from lepton energy moments and $q^2$ moments, assuming $\delta \langle\mathcal{M}\rangle =\pm 0.1$.}
	\label{fig:SLT}
\end{figure}

From our results, we observe that for all moments the  contribution proportional to $C_T^2$ is sizable compared to $\xi_{\rm SM}$.
Especially for the third $E_\ell$ and $q^2$ moments, tensor contributions can be as large as
ten times the SM prediction or more (for order one coefficients). Therefore, a moment analysis is expected to be able to strongly constrain such contributions. It is also interesting to consider the case of contributions from both $C_{S_L}$ and $C_T$, because due to RGE running (see e.g.~\cite{Gonzalez-Alonso:2017iyc,Hu:2018veh}), tensor interactions always generate left-handed scalar interactions. We note that $q^2$ moments are only sensitive to the quadratic contributions, while lepton and hadronic mass moments are also sensitive to interference. Assuming real couplings and $C_{S_L}>C_T$ (see discussion in \cite{Jung:2018lfu}), we observe that the $q^2$ moments mainly constrain $C_{S_L}$, while the lepton moments constrain the tensor part. Clearly, the situation for the inclusive decay is not as straightforward as for the exclusive case, because our current ``SM prediction'' depends on the input of the HQE elements that are extracted from data. Nevertheless, we can visualize and investigate the potential NP bounds for different scenarios by assuming that the SM prediction is known (namely $\xi_{\rm SM}$). We then define
\begin{align}
    \delta \braket{\mathcal{M}} \equiv %\left |
    \frac{\braket{\mathcal{M}} -\braket{\mathcal{M}}_\text{SM}}{\braket{\mathcal{M}}_\text{SM}} %\right|
    \,
    \label{eq:NP-SM-impact}
\end{align}
where $\mathcal{M}_\text{SM} = \xi_{\rm SM}$ for the specific moment under consideration. Considering then a $10\%$ measurement of the moments, i.e.\ $\delta\braket{\mathcal{M}}=\pm 0.1$, leads to a constraint on the NP parameters. Specifically, for the $S_L-T$ contributions we obtain 
\begin{equation}\label{eq:npcons}
 -0.1<   |C_{S_L}|^2 \hat{\xi}_{\rm NP}^{\langle S_L, S_L\rangle} +  |C_{T}|^2 \hat{\xi}_{\rm NP}^{\langle T, T\rangle} + {\rm Re}(C_{S_L}C_T^*)  \hat{\xi}_{\rm NP}^{\langle S_L, T\rangle}  <0.1 \ ,
\end{equation}
where 
\begin{equation}
    \hat{\xi}_i \equiv \frac{\xi_i}{\xi_{\rm SM}} \ ,
\end{equation}
and the $\xi_i$ can be found in Appendix~\ref{sec:appxi} for the different moments and NP scenarios. In order to illustrate the effects, we use these $\xi$'s, which are re-expanded in the Wilson coefficients. The constraints obtained from \eqref{eq:npcons} are illustrated in Fig.~\ref{fig:SLT}. Interestingly, we see that the different moments give complementary bounds on NP, similar as the $B\to D$ versus $B\to D^*$ constraints in the exclusive case (for the latter see \cite{Jung:2018lfu}). 

\begin{figure}
	\centering
	{\includegraphics[width=0.85\textwidth]{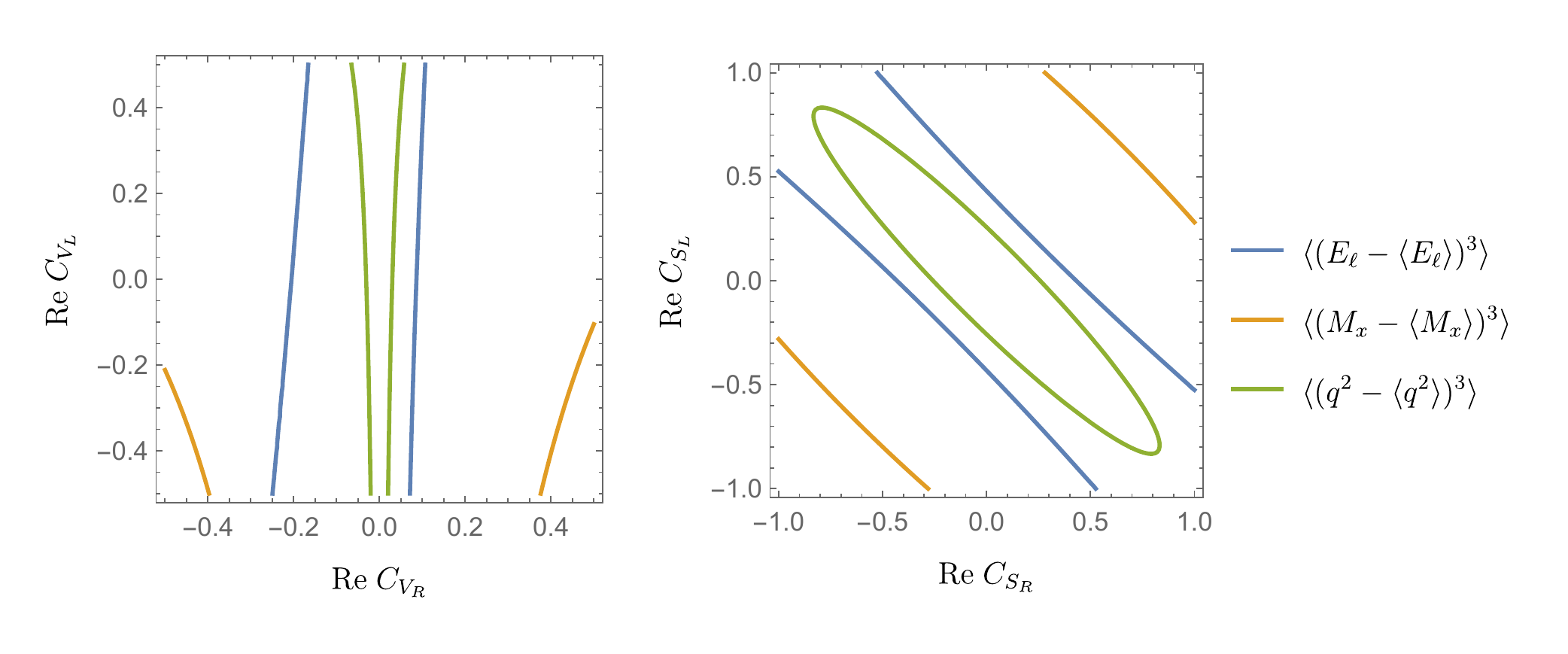}}
  \caption{Illustration of the possible bounds on (left) $C_{V_L}$ versus $C_{V_R}$ and (right) $C_{S_L}$ versus $C_{S_R}$ assuming a $10\%$ SM measurement. }
	\label{fig:othernp}
\end{figure}

Similarly, in Fig.~\ref{fig:othernp}, we illustrate the possible bounds on $C_{V_L}$ and $C_{V_R}$ (left) and $C_{S_L}$ and $C_{S_R}$ (right). In these cases, we see that the $M_X$ moments give much weaker constraints than the lepton energy and $q^2$ moments. We should stress that the uncertainties on the $M_X$ moments are in general also larger as they are more sensitive to higher-order HQE corrections. Comparing with the exclusive constraints on $C_{S_L}$ versus $C_{S_R}$ in \cite{Jung:2018lfu}, we observe that such a SM measurement would constrain NP along the $C_{S_L}=-C_{S_R}$ plane, similar as the $B\to D$ exclusive mode, while $B\to D^*$ gives constraints orthogonal to that.

Finally, we note that the $C_{S_{L,R}} C_{T}$ coefficient vanishes for $q^2$ moments because
the differential rate has only a parity-odd contribution while $q^2$ moments with a cut on $q^2$ are parity even observables. For lepton energy and hadronic mass moments, the contribution proportional to $C_{S_{R}} C_{T}$  is non-zero only due to power corrections. Therefore, the sensitivity to these types of NP is limited. 

\subsection{Illustration for specific NP scenarios}

\begin{table}[t]
\begin{center}
\scalebox{0.9}{
\begin{tabular}{ |c | c| c| c| c | c| }
\hline
NP Scenarios &  $C_{V_L}$ & $C_{V_R}$ & $C_{S_R}$ & $C_{S_L}$ & $C_{T}$\\
\hline
I &  0 & 0 &  1 & 1 & 0 %I &  0 & 0 & -2 $C_{S_L}$ & 1.5 & 0
\\
\hline
II & 0 & 0 & 0 & -1 & 0.5
\\ 
\hline
III & -1 & 0.5 & 0 & 0 & 0
\\
\hline
\end{tabular}}
\end{center}
\caption{Three NP scenarios that we consider to visualize the effect of the NP parameters in the moments. 
All Wilson coefficients are defined at the scale $\mu=m_b$.}
\label{table:NPScen}
\end{table}
To visualize the effect of possible NP in the moments of the $B\to X_c \ell \bar{\nu}_\ell$ spectrum as a function of the lepton energy cut (or $q^2$ cut), we consider three NP scenarios specified in Table \ref{table:NPScen} allowing for either new scalar interactions (Scen.~I), new tensor and scalar interactions (Scen.~II) and new vector interactions (Scen.~III). 
These scenarios are just to illustrate how the NP contributions depend on the cut and to the SM uncertainty. We stress that these scenarios may not be realistic in light of current data on exclusive $B\to D^{(*)}$ decays, were the same NP operators would contribute. Specifically Scenario II, where we allow for a rather large tensor contribution, may be already excluded by the exclusive decays (see \cite{Jung:2018lfu}). 
For the scalar contributions, we pick $C_{S_R}=C_{S_L}$, based on Fig.~\ref{fig:othernp} as we see that this would give a large effect on the spectrum. Finally, as here we consider rather large Wilson coefficients we do not re-expand the expression for the moments in the Wilson coefficients. We observe in Figs.~\ref{fig:lepton-moment}, \ref{fig:MxMoment} and \ref{fig:q2Moments} that the prediction for all central moments are modified by the presence of NP contributions, but that the cut-dependence remains similar as that of the SM prediction. For all cases, we observe that the third central moment is most sensitive to NP effects. 

\subsubsection*{Electron energy moments:}
\begin{figure}
	\centering
	\subfloat{\includegraphics[width=0.5\textwidth]{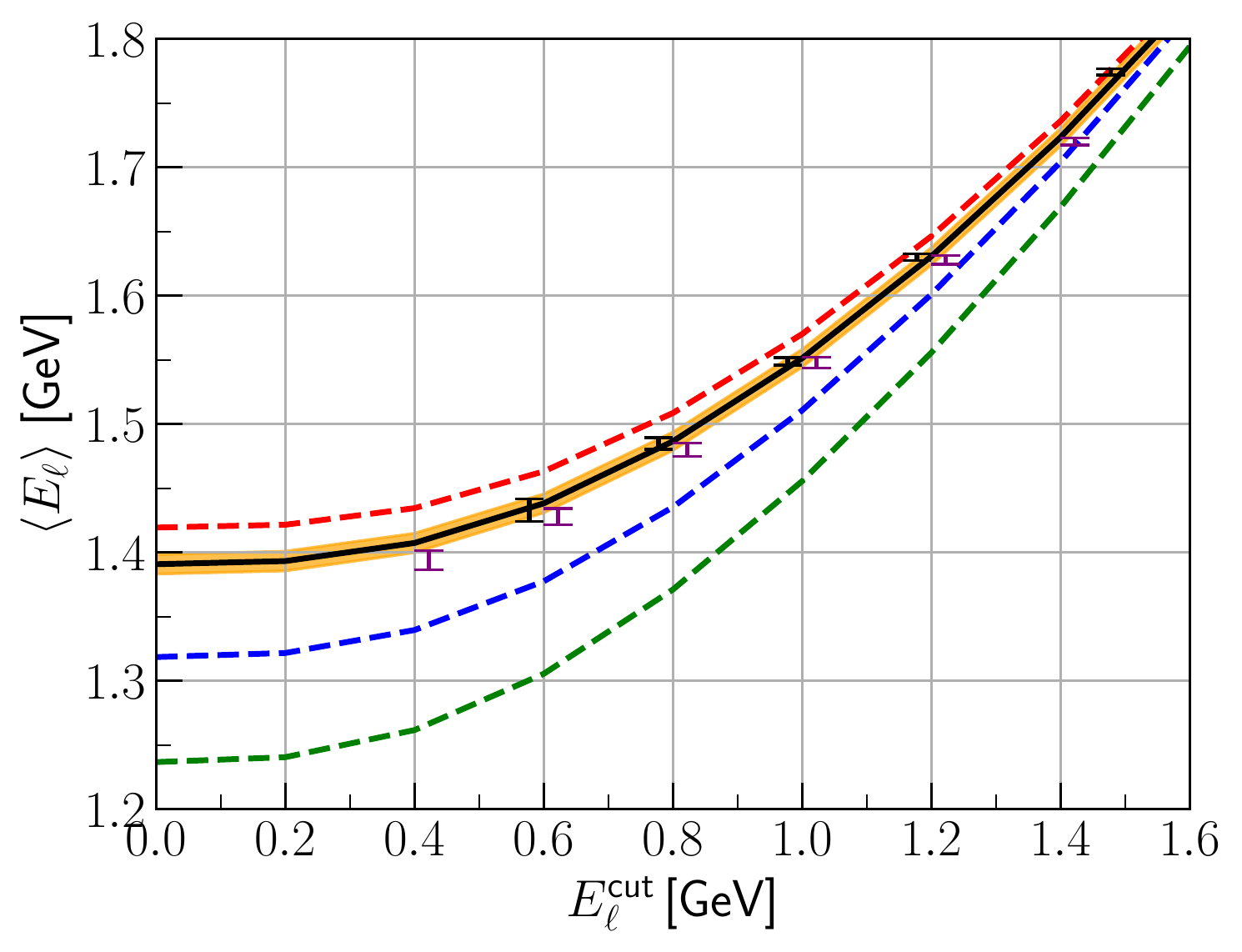}}
	\subfloat{\includegraphics[width=0.5\textwidth]{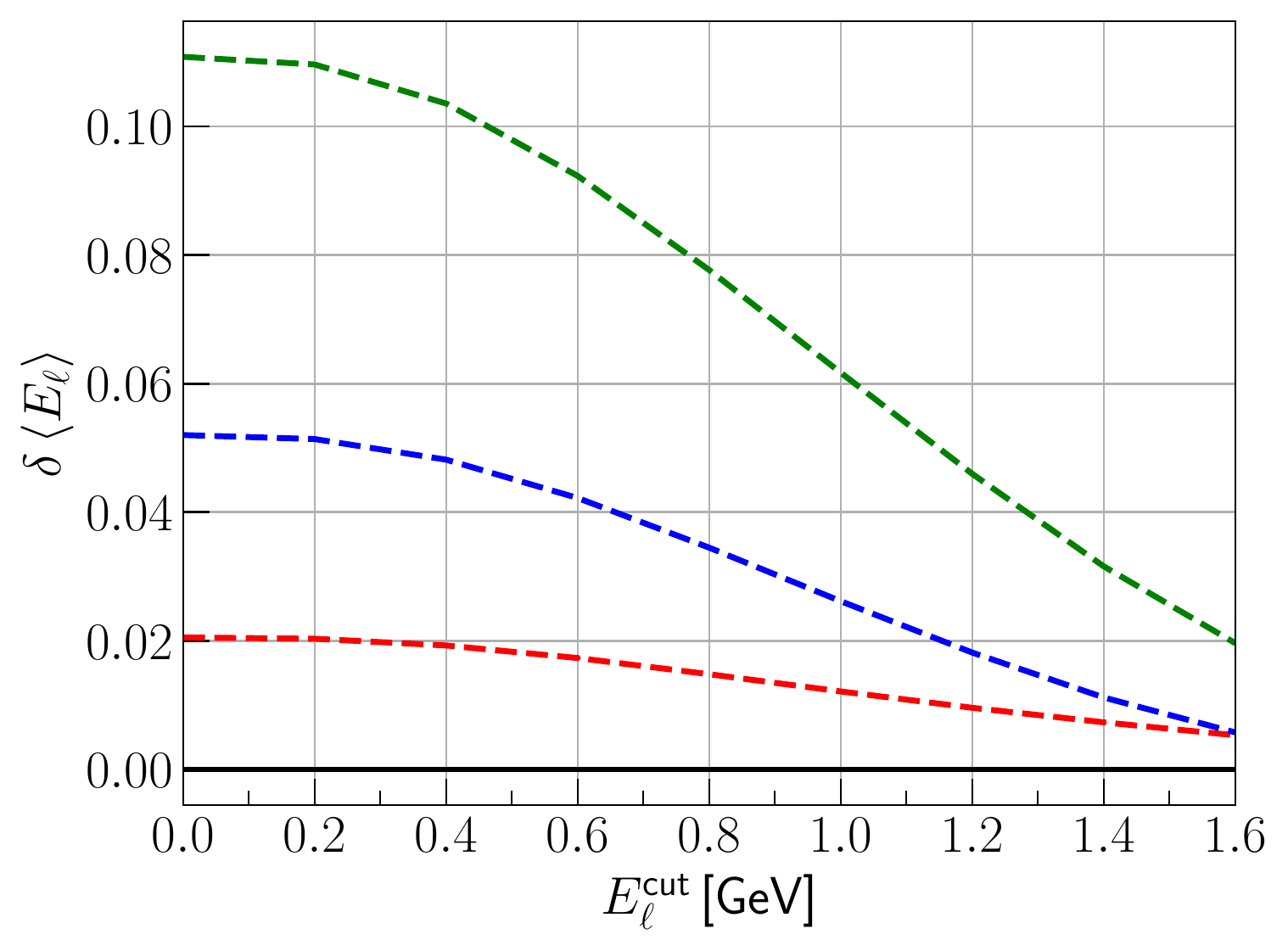}} \\
  	\subfloat{\includegraphics[width=0.5\textwidth]{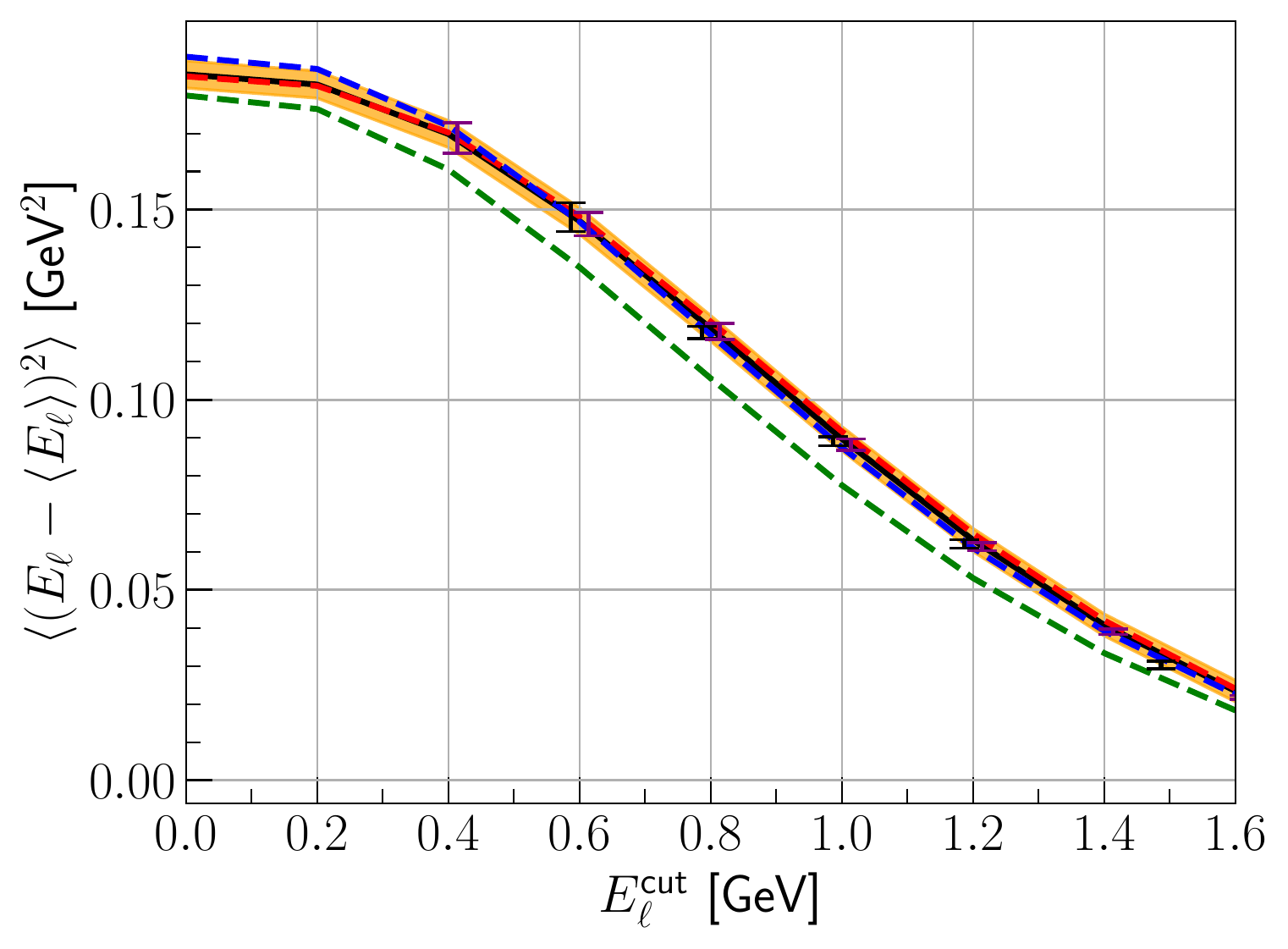}}
  	\subfloat{\includegraphics[width=0.5\textwidth]{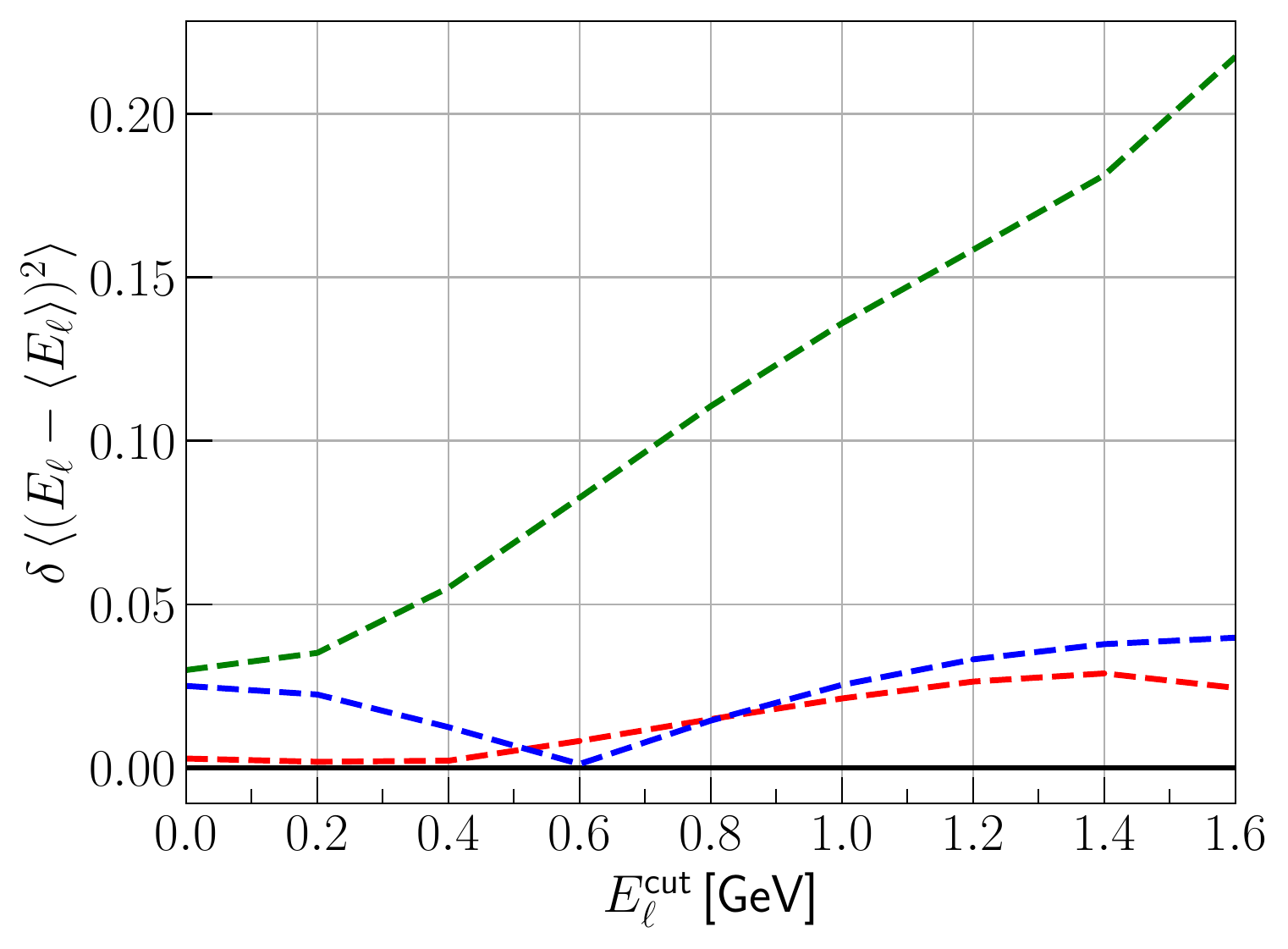}} \\
  	\subfloat{\includegraphics[width=0.5\textwidth]{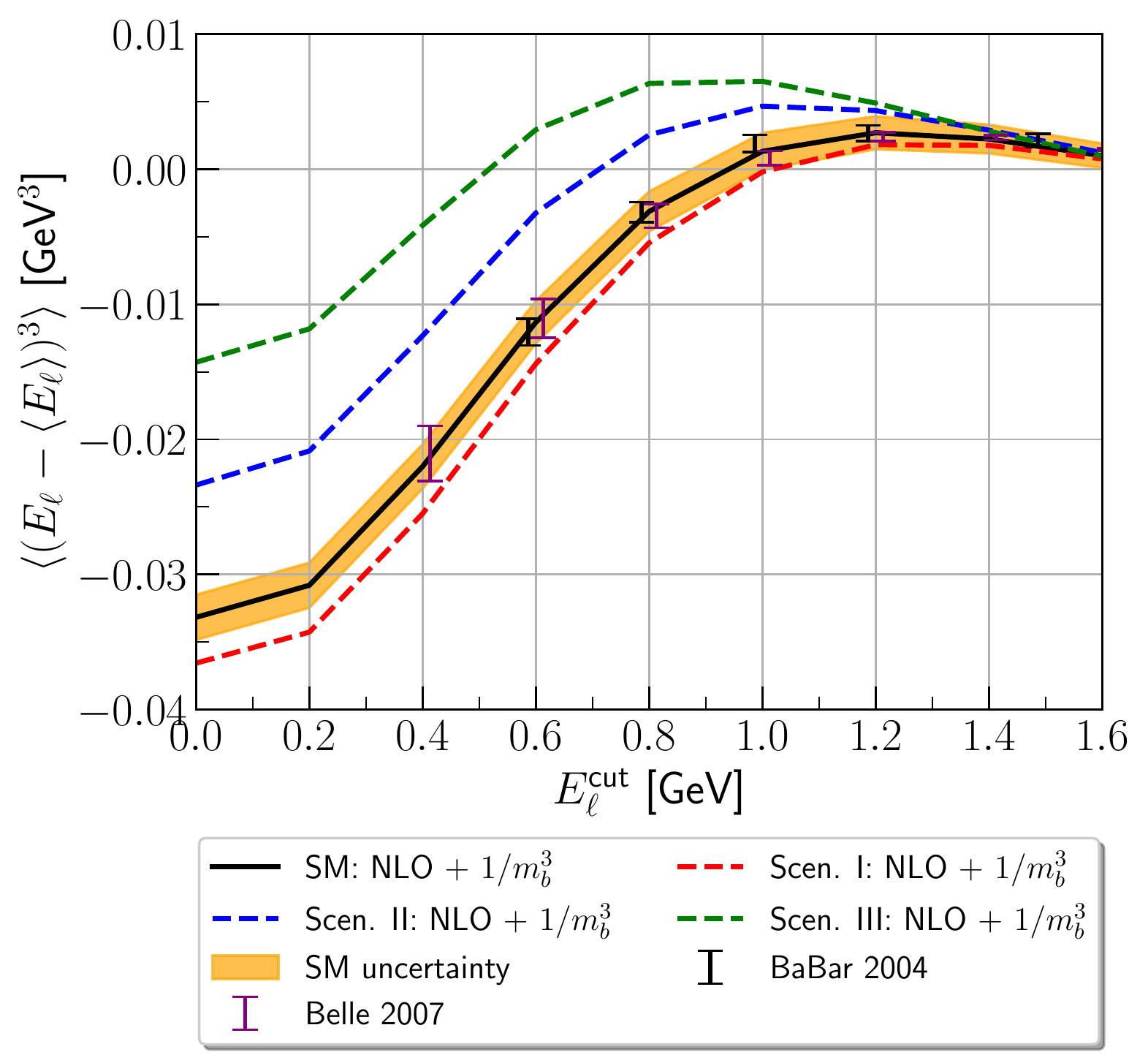}}
  	\subfloat{\includegraphics[width=0.5\textwidth]{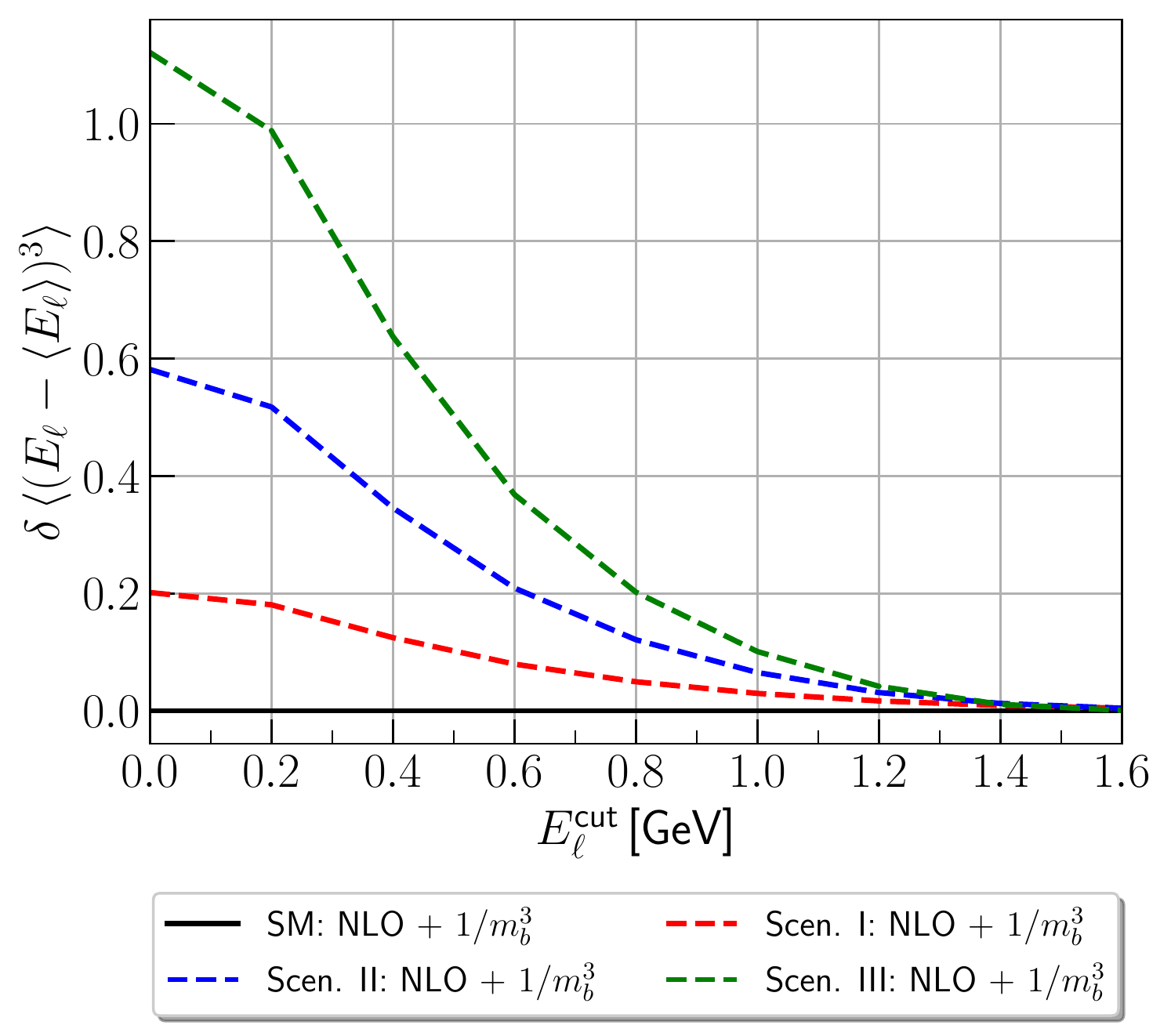}}\\
  \caption{Lepton energy moments for the $B \to X_c \ell \bar{\nu}_\ell$ decay for the different NP
  scenarios (see Tab.~\ref{table:NPScen}). 
  The experimental results of BaBar is taken from \cite{BaBar:2004bij} and Belle from \cite{Belle:2006kgy}. }
	\label{fig:lepton-moment}
\end{figure}

Figure~\ref{fig:lepton-moment} shows the lepton energy moments as a function of the lepton energy cut for the SM and the three NP scenarios. 
In order to qualitatively understand the sensitivity on possible NP effects, 
we show in these plots the experimental results from Belle \cite{Belle:2006kgy} and BaBar \cite{BaBar:2004bij}. 
On the right-hand side, we show the impact of the NP scenarios by showing the absolute value of $ \delta \braket{\mathcal{M}} $ defined in \eqref{eq:NP-SM-impact}. 

For simplicity, we only show an uncertainty band for the SM prediction obtained by varying the inputs in Table~\ref{table:input} within their $1\sigma$ ranges.  
To account for missing $\alpha_s$ corrections, we vary the scale of $\alpha_s(\mu)$ in the range $m_b/2 < \mu < 2 \, m_b$. 
We observe for electron energy moments, Scen.~I is rather close to the SM, while Scen.~II and III cause a shift much larger than the SM uncertainty.
These lepton energy moments therefore seem rather sensitive to NP effects and it would be potentially able to constrain NP via a full global analysis of these moments.
Note also that the contribution from power corrections are in general small for this kind of moments, reducing the dependence on the value of the HQE parameters.

\subsubsection*{\boldmath $M_X^2$ moments:}
\begin{figure}
	\centering
	\subfloat{\includegraphics[width=0.5\textwidth]{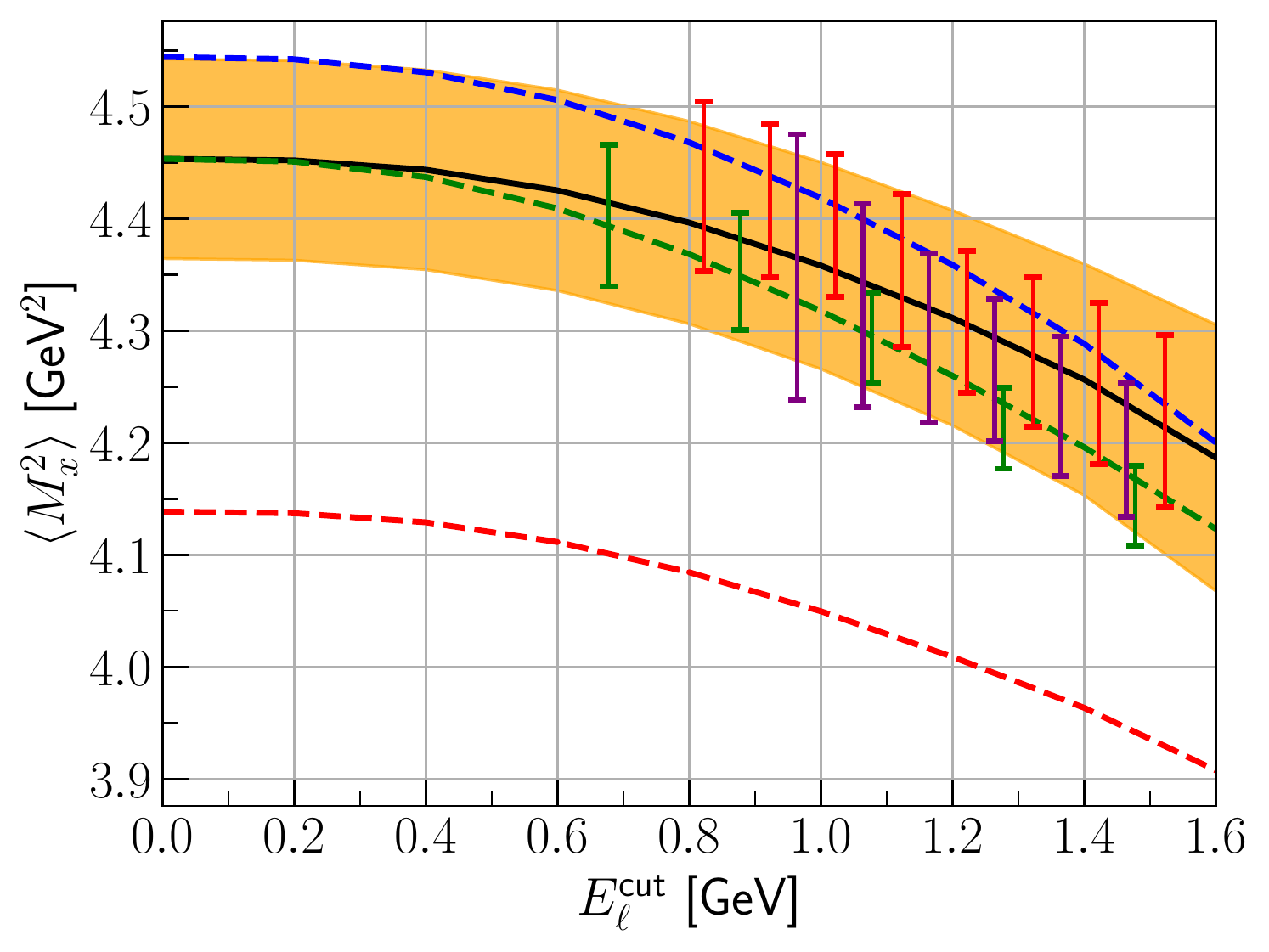}} 
	\subfloat{\includegraphics[width=0.5\textwidth]{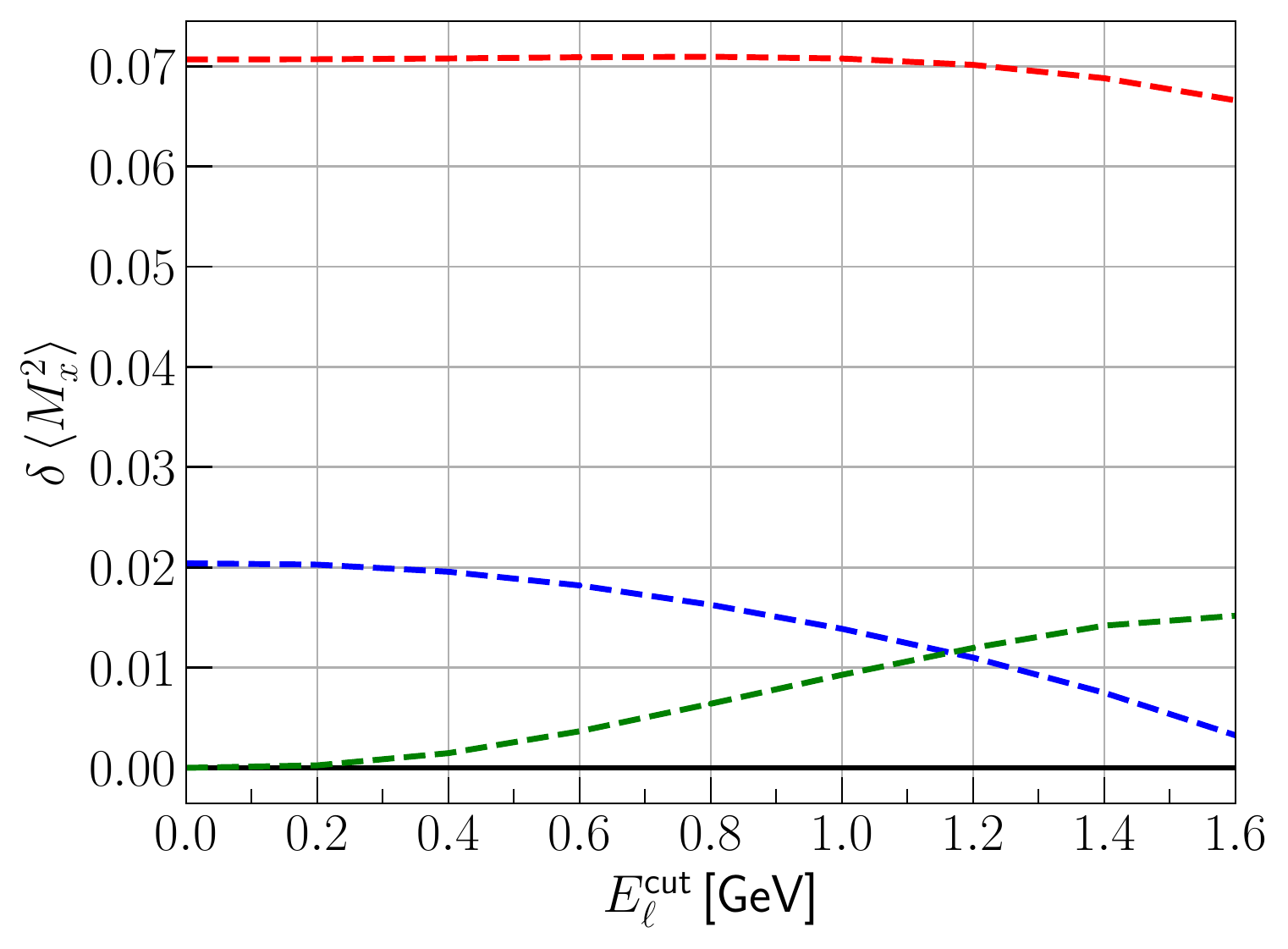}}\\ 
  	\subfloat{\includegraphics[width=0.5\textwidth]{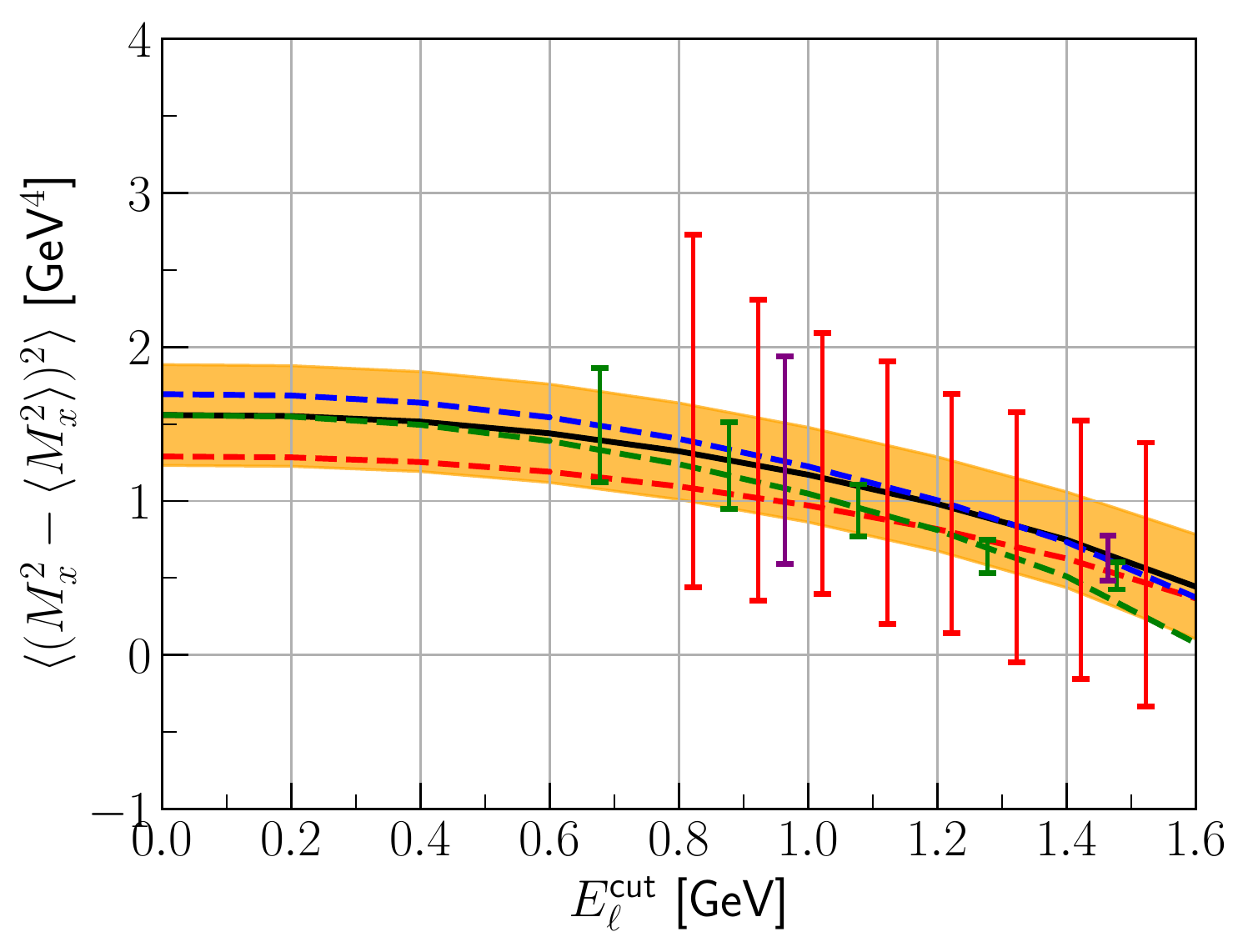}} 
  	\subfloat{\includegraphics[width=0.5\textwidth]{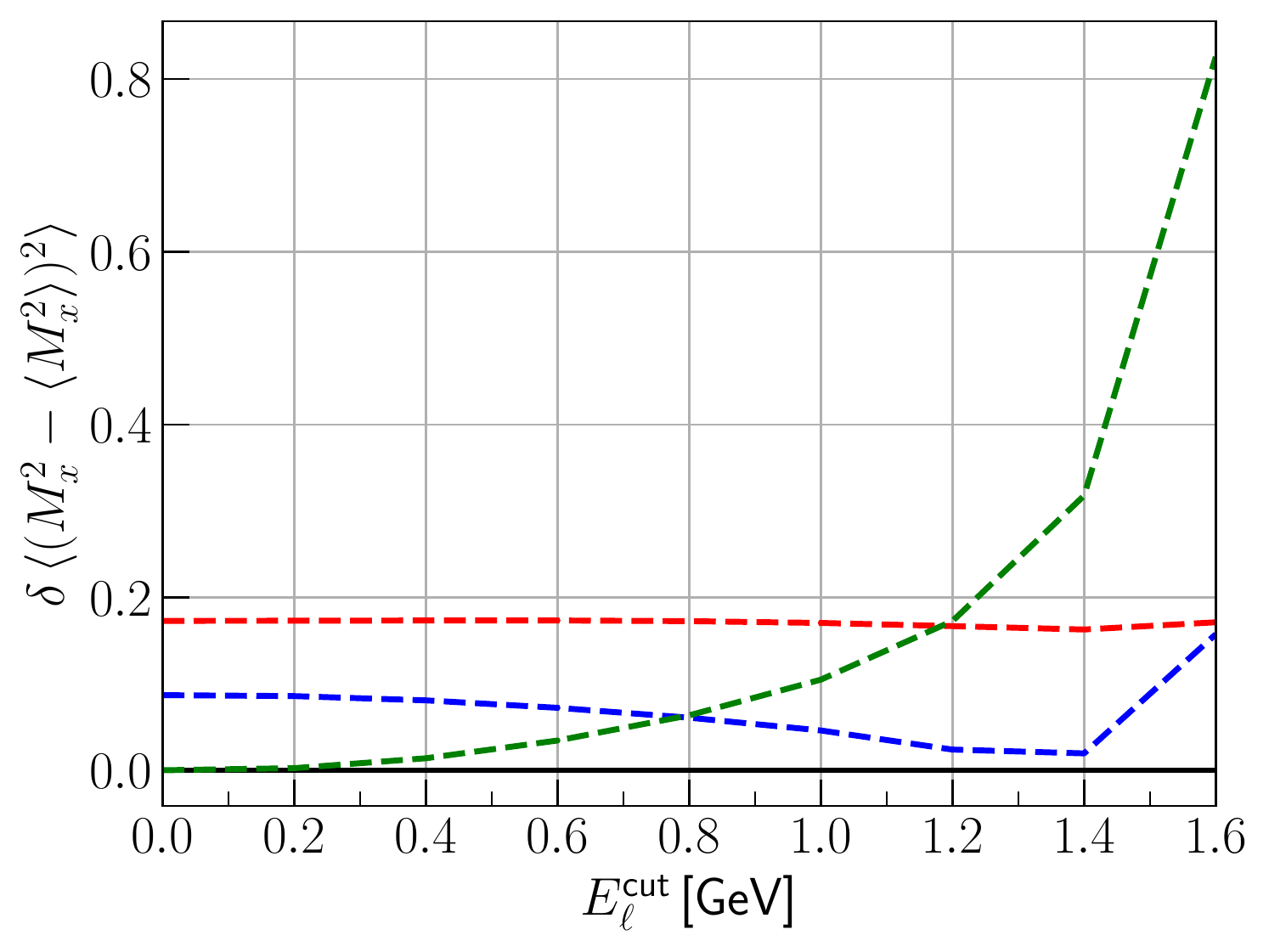}} \\ 
  	\subfloat{\includegraphics[width=0.5\textwidth]{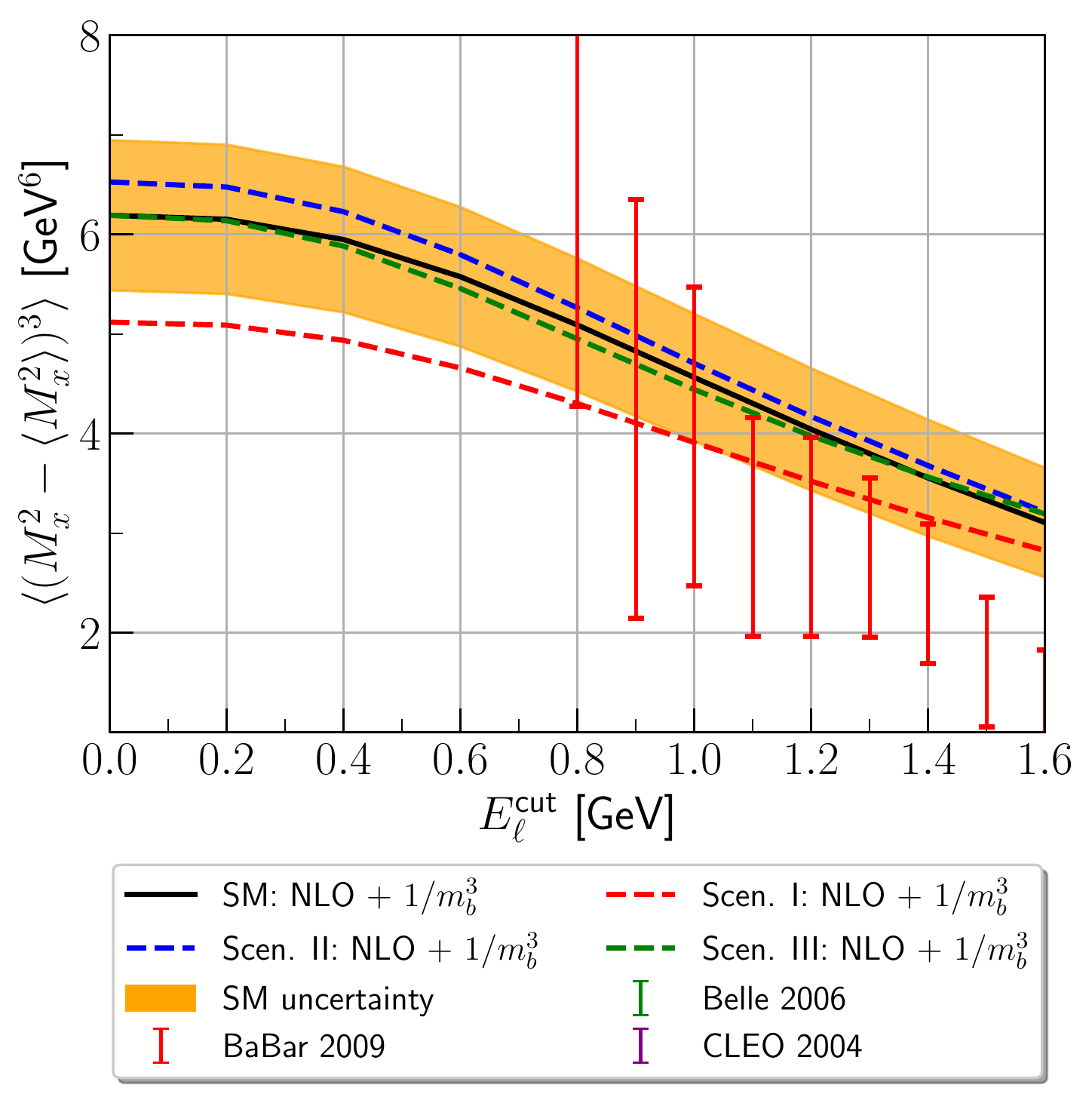}}
  	\subfloat{\includegraphics[width=0.52\textwidth]{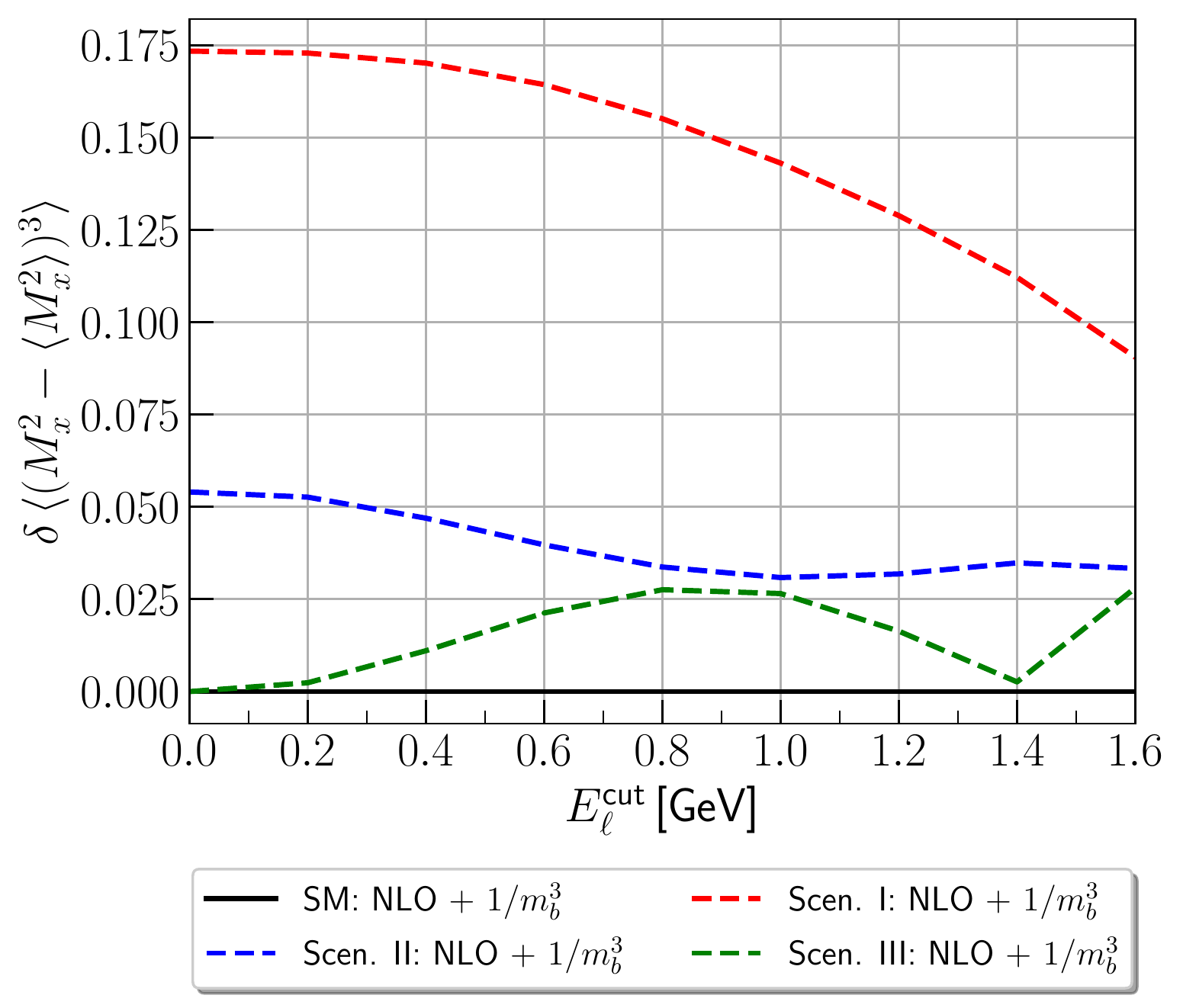}}
  \caption{Hadronic invariant mass moments for the $B \to X_c \ell \bar{\nu}_\ell$ decay 
  different NP  scenarios (see Tab.~\ref{table:NPScen}).
  The experimental values of BaBar is taken from \cite{BaBar:2009zpz}, CLEO \cite{CLEO:2004bqt}, Belle \cite{Belle:2006jtu}.}
	\label{fig:MxMoment}
\end{figure}

Results for the hadronic invariant mass moments are shown in Fig.~\ref{fig:MxMoment}. We observe that these moments are sensitive to new scalar couplings, as Scen.~I shows the largest deviation from the SM prediction. On the other hand, both Scen.~II and III lie within the uncertainty of the SM error band, which is rather large. This happens because for the $M_X$ moments the contribution from power corrections is very important and the $\alpha_s$ corrections are much larger compared 
to the partonic LO. The dependence of the $M_X$ moments on the scale of $\alpha_s$ is therefore much 
larger compared for instance to the electron energy moments and so prevents a precise SM determination of these kind of observables. We thus conclude that currently it will be rather challenging to only use the $M_X$-moments to distinguish between the SM and NP scenarios.

For the experimental data points we use the results of CLEO~\cite{CLEO:2004bqt}, Belle~\cite{Belle:2006jtu} and BaBar~\cite{BaBar:2009zpz}. The latter does not provide the central moments but only $\braket{(M_X^2)^{i=1,2,3}}$. We have calculated the central moments using \eqref{eq:central-moments}. We do not show the recent results of Belle II~\cite{Belle-II:2020oxx} since the uncertainties are still rather large. 

\subsubsection*{\boldmath $q^2$ moments:}
\begin{figure}
	\centering
  	\subfloat{\includegraphics[width=0.5\textwidth]{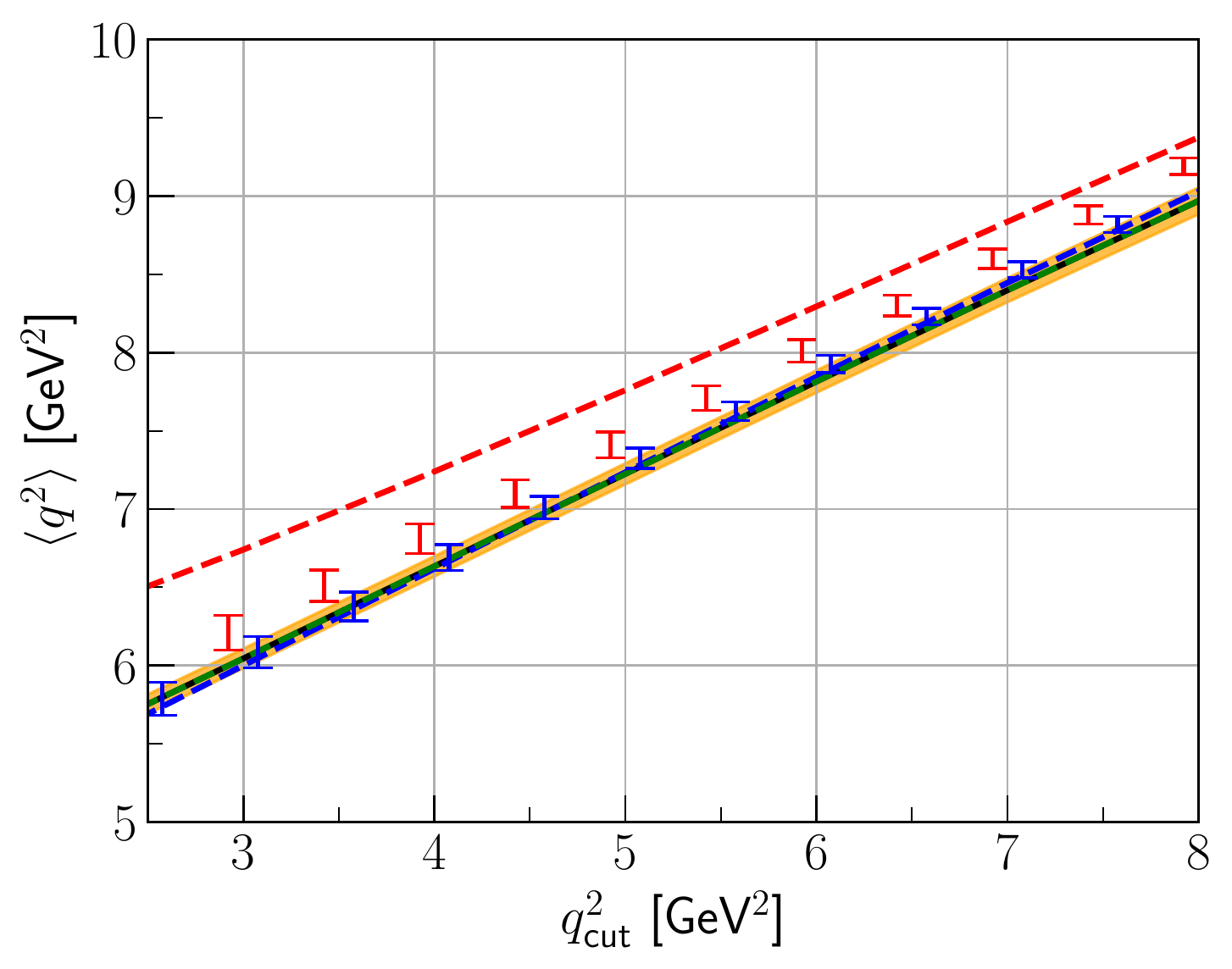}}
  	\subfloat{\includegraphics[width=0.5\textwidth]{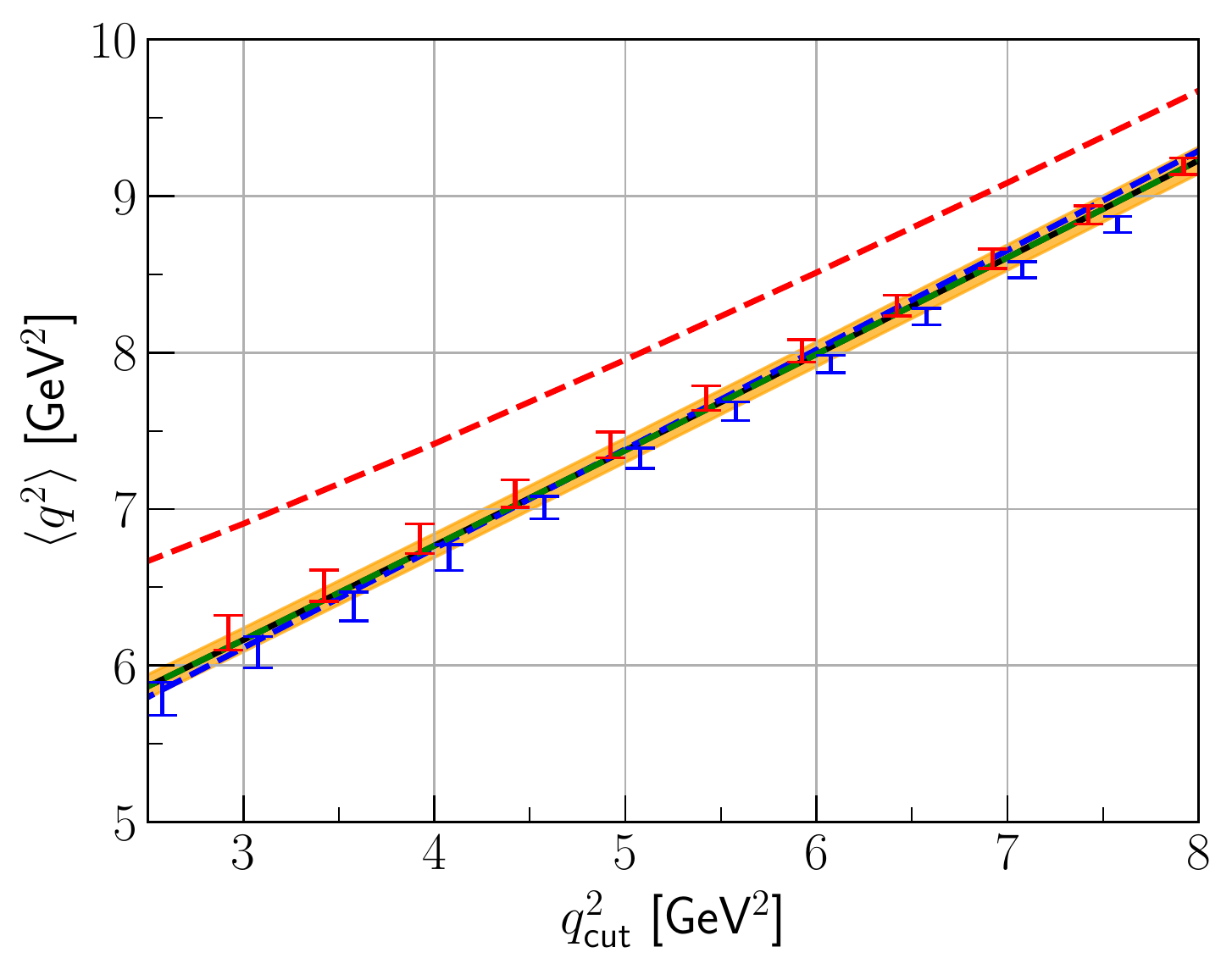}}\\
	\subfloat{\includegraphics[width=0.5\textwidth]{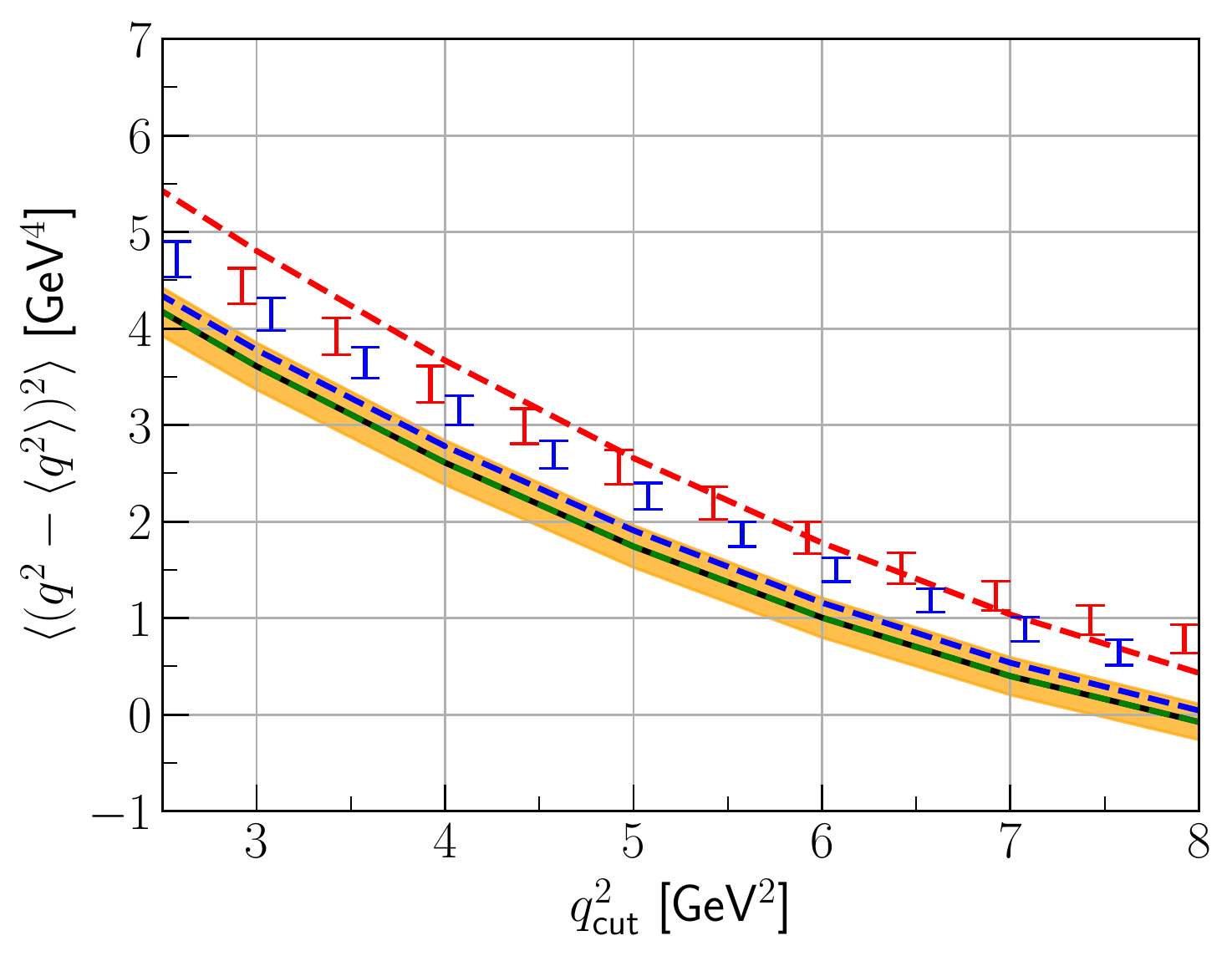}}
	\subfloat{\includegraphics[width=0.5\textwidth]{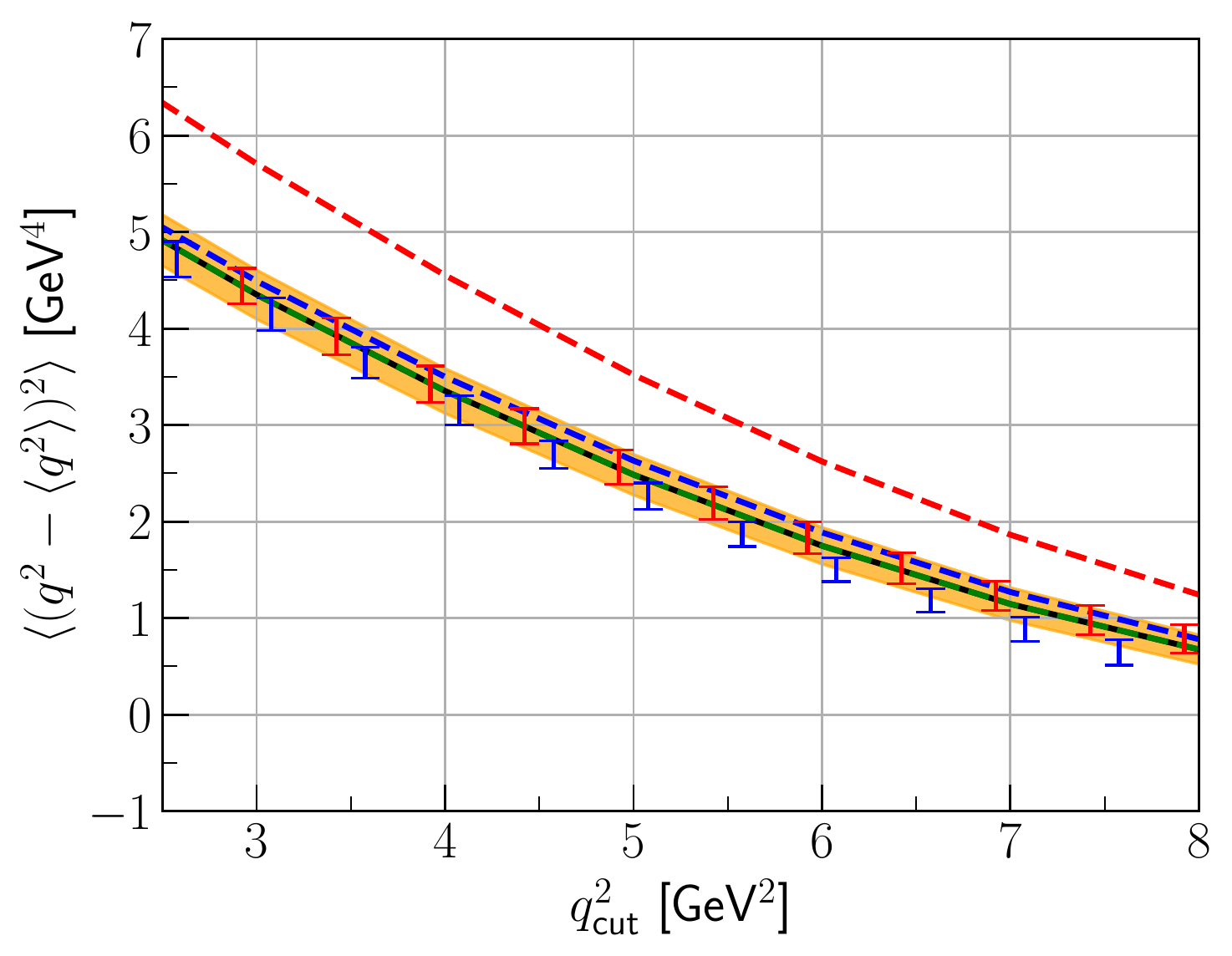}}\\
  	\subfloat{\includegraphics[width=0.5\textwidth]{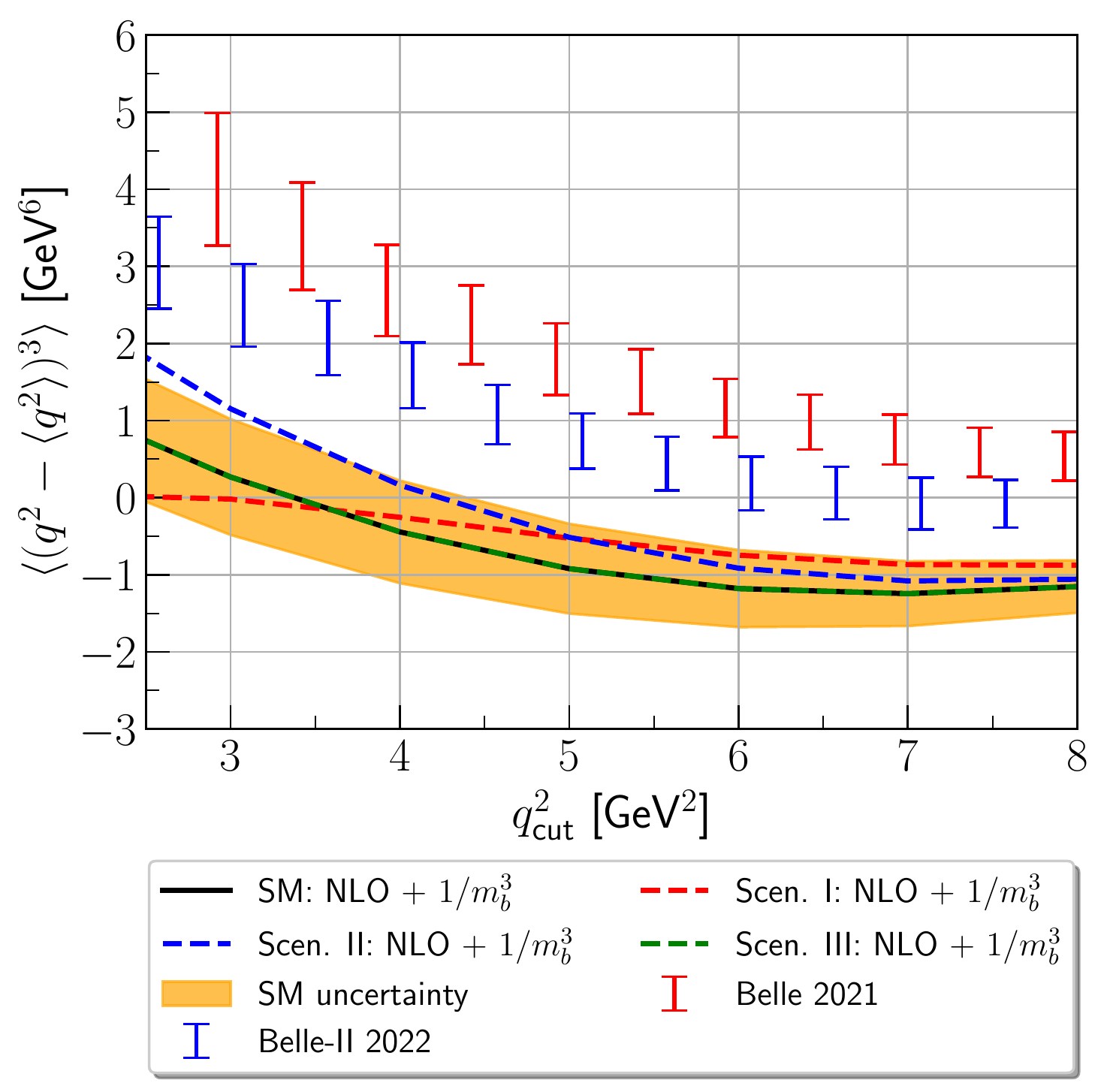}}
  	\subfloat{\includegraphics[width=0.51\textwidth]{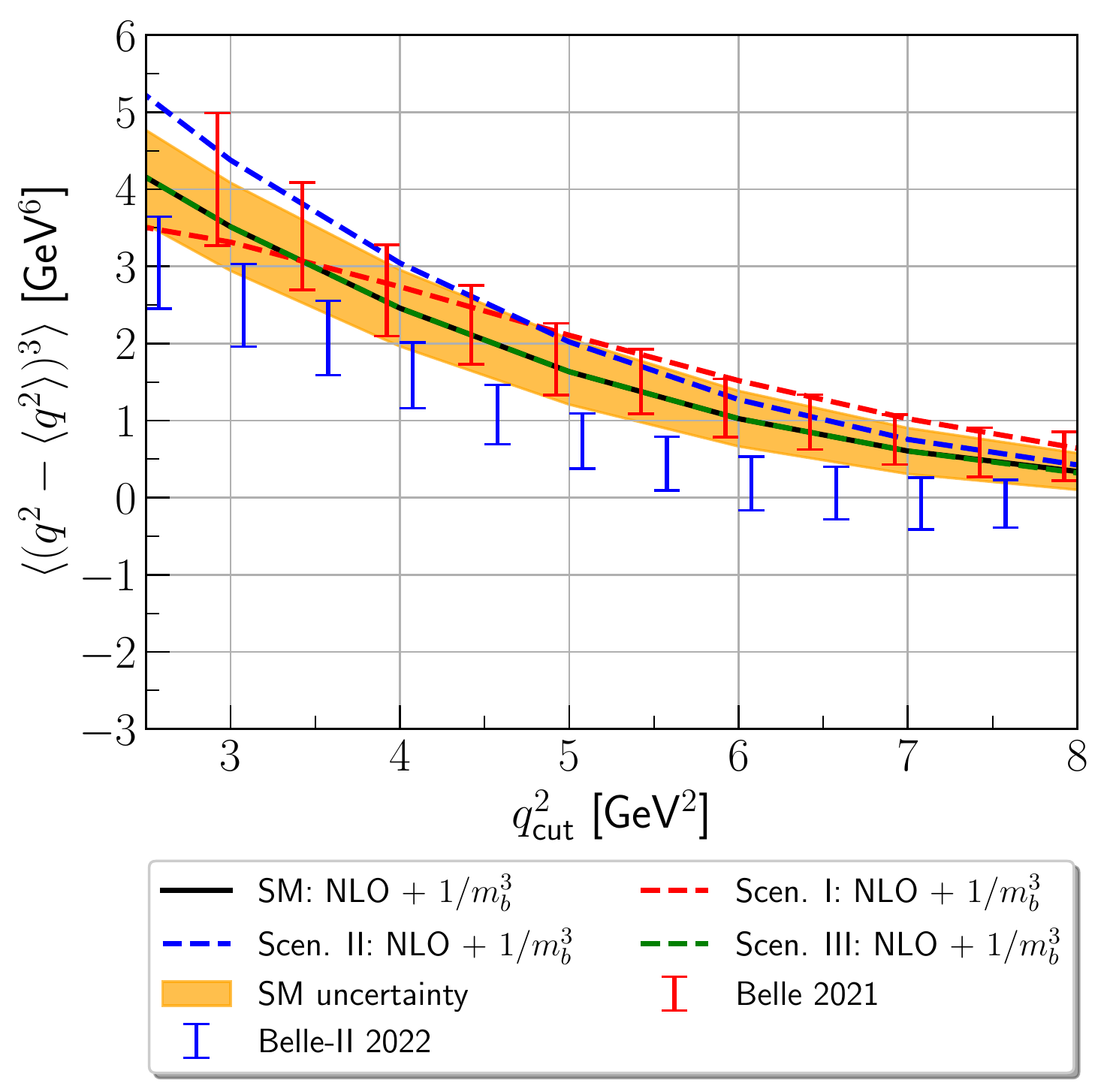}}
  	\caption{Dilepton invariant mass moments ($q^2$) for the $B \to X_c \ell \bar{\nu}_\ell$ decay in comparison with Belle \cite{Belle:2021idw} and Belle II data \cite{The:2022cbm}. (Left) Using the inputs in Table~\ref{table:input} from \cite{Bordone:2021oof}(Right) using the inputs from \cite{Bernlochner:2022ucr}.}
	\label{fig:q2Moments}
\end{figure}

For the $q^2$ moments, we consider the SM and NP predictions at different values of the $q^2$ cut shown in Figure~\ref{fig:q2Moments}. For the plots on the left-hand side, we used the HQE parameters from Table~\ref{table:input} from \cite{Bordone:2021oof}. 
Comparing with the experimental data points of Belle~\cite{Belle:2021idw} and Belle II~\cite{The:2022cbm}, we find large deviations. Interestingly, these deviations cannot be accommodated by the three NP scenarios we consider.
As mentioned before, in \cite{Bernlochner:2022ucr}, where these data were used to extract the HQE parameters and $V_{cb}$, a value of $\rho_D^3$ incompatible with that in Table~\ref{table:input} was found. 
The mismatch in Fig.~\ref{fig:q2Moments} is a consequence of this: the $q^2$ data pull $\rho_D^3$ to much smaller value. To illustrate this, we show on the right-hand side of Fig.~\ref{fig:q2Moments} the SM predictions using the HQE parameters obtained in \cite{Bernlochner:2022ucr}. We observe good agreement with the data points. In addition, the uncertainty of the SM prediction is rather large, reflecting that these moments are more sensitive to the power corrections than the lepton energy moments. This was already observed in \cite{Fael:2022frj}. Note that the $\xi$ coefficients in Appendix~\ref{sec:appxi} are obtained using Table~\ref{table:input}. As the goal of these scenarios is merely to demonstrate the effect of different NP parameters, we do not present $\xi$'s using the HQE parameters from \cite{Bernlochner:2022ucr}. We observed the $q^2$ moments are most sensitive to Scen.~I, while Scen.~III has basically no effect. This is because for this scenario there is a cancellation between the Wilson coefficients, rendering the effect almost unobservable. For smaller values of $C_{V_L}$, there is an effect on the moments and in fact the $q^2$ moments can put rather strong constraints as seen in Fig.~\ref{fig:othernp}.

\subsection{Lepton Flavor Universality Ratios}
In order to study NP in Lepton Flavor Universality Ratios of light leptons, we give the analytic expression for the total rate in Appendix~\ref{app:nptot}.  

\begin{table}[t]
\begin{center}
\renewcommand{\arraystretch}{1.2} 
\begin{tabular}{c c }
\toprule
 & $\mathcal{B}(B \to X_c \ell \bar{\nu})$ in \%   \\
\midrule
$\xi_{\text{SM}}$ & $12.983 |_\text{LO} -0.962 |_\text{pow} - \left( \frac{\alpha_s}{\pi} \right) 16.101 $  \\
$\xi^{\braket{V_R ,V_R}}_{\text{NP}}$ & $12.983 |_\text{LO} -0.962 |_\text{pow} - \left( \frac{\alpha_s}{\pi} \right) 16.101 $  \\
$\xi^{\braket{S_L ,S_L}}_{\text{NP}}$ & $ 3.245|_\text{LO} +0.067|_\text{pow} + \left( \frac{\alpha_s}{\pi} \right) 2.783 $  \\
$\xi^{\braket{S_R ,S_R}}_{\text{NP}}$ & $ 3.245|_\text{LO} +0.067|_\text{pow} + \left( \frac{\alpha_s}{\pi} \right) 2.783 $  \\
$\xi^{\braket{T ,T}}_{\text{NP}}$ & $155.802 |_\text{LO} -16.493|_\text{pow}  -\left( \frac{\alpha_s}{\pi} \right) 163.665 $ \\
$\xi^{\braket{V_L ,V_R}}_{\text{NP}}$ & $-8.453|_\text{LO} +1.332|_\text{pow} +\left( \frac{\alpha_s}{\pi} \right) 13.375 $  \\
$\xi^{\braket{S_L ,S_R}}_{\text{NP}}$ & $ 4.226 |_\text{LO} + 0.380|_\text{pow} + \left( \frac{\alpha_s}{\pi} \right) 4.550 $  \\
$\xi^{\braket{S_L ,T}}_{\text{NP}}$ & 0  \\
$\xi^{\braket{S_R , T}}_{\text{NP}}$ & 0  \\
\toprule
\end{tabular}
\caption{Numerical values of the parameters for the branching ratio without lepton energy cut for fixed $B$ meson lifetime. }
\label{tab:Br}
\end{center}
\end{table}

For completeness, we also give the numerical coefficients including NLO correction. Writing the branching ratio in terms of $\xi_{i}$ as in \eqref{eq::moments}, with the only difference that the $\xi_{\rm SM}$ term gets multiplied with $|1+C_{V_L}|^2$, we find the coefficients listed in  Table~\ref{tab:Br}. We used the inputs in Table~\ref{table:input} and a fixed value for the $B$ meson lifetime $\tau_B = 1.579$ ps \cite{10.1093/ptep/ptaa104} and $|V_{cb}| = (42.16 \pm 0.51) \cdot 10^{-3}$ \cite{Bordone:2021oof}. However, we note that NP would also affect the total lifetime of the $B$ meson. 

The expressions in the Appendix and our numerical results can be used to study ratios of electron versus muon rates under the assumption of lepton-flavour universality violating new physics. Note that a lepton-flavour universal and diagonal NP effect in $C_L$
can in principle be absorbed by a shift in $V_{cb}$. 
Recently, the SM predictions for lepton-flavour universality ratios were studied \cite{Rahimi:2022vlv}. Because the current data (see for example the $q^2$ moments split up for electron and muon contributions in \cite{Belle:2021idw}) do not indicate any deviation from lepton universality in the charged light modes, we do not study these effects here further. 

\subsection{HQE parameters versus NP}

The HQE parameters are extracted from moments of the $b\to c$ spectrum under the assumption of the SM. However, it can be that NP mimics the effect of the HQE parameters shifting the spectrum up or down. In fact, Fig.~\ref{fig:q2Moments} shows that shifting $\rho_D^3$ seems to be able to mimic the effect that NP may have on the spectrum. It would therefore be interesting to perform a full analysis of the moments including NP. Such an analysis lies beyond the scope of the current paper. However, we can illustrate the possible effect with a simplified toy fit. To this extend, we generate pseudo data points for the three NP scenarios in Table~\ref{table:NPScen} for lepton energy and hadronic invariant mass moments at different lepton energy cuts as well as $q^2$ moments with $q^2_{\text{cut}}$. For this, we use the HQE parameters in Table \ref{table:input}. We generate 9 data points per scenario: the first, second and third central moments with $E_\ell^{\rm cut} = 1.0$ GeV for the lepton energy and hadronic invariant mass moments and with $q^2_{\text{cut}} = 4$ GeV$^2$ for $q^2$ moments. For the uncertainty on these points,  we vary the contribution of $\rho_D^3$ by $30\%$, $\mu_G^2$ by $20\%$ and $\alpha_s$ between its value at $\mu=m_b/2$ and $\mu=m_b$, based on \cite{Bordone:2021oof, Bernlochner:2022ucr}. As this render the uncertainty for the lepton energy moments rather small, we add an additional uncertainty based on the current experimental uncertainty. In addition, we also include the current experimental uncertainty for the third $q^2$ and $M_X$ moments as these are rather large. 

In principle, these pseudo data points can then be used to fit for the HQE parameters $\mu_G^2, \mu_\pi^2,\rho_{LS}^3, \rho_D^3$ using the SM expressions. In this way, our toy fit mimics a situation that may happen in reality: i.e.\ NP is present but the extraction of HQE parameters is done assuming the SM. We observe that for the three NP scenarios in Table~\ref{table:NPScen}, our simple toy fit yields large $\chi^2$. The reason for this is that it is challenging to accommodate the third moments, which are sensitive to NP, and first lepton energy moments, which drives the fit due to its small uncertainty, at the same time. Turning the argumentation around this may indicate that a full simultaneous fit of the HQE parameters and NP parameters would give rather good constraints on NP. In this endeavour, it seems crucial to improve the experimental inputs especially on the third moments. 

Finally, we may also consider a more realistic scenario taken from the analysis of \cite{Jung:2018lfu}: $C_T=0.05$ and $C_{S_L}=-0.5$. Assuming no correlations between the pseudo data points, we obtain a $\chi^2/d.o.f. \simeq 2.4$ and
\begin{equation}
    \mu_G^2|_{\rm toy} = 0.40\; {\rm GeV}^2, \;\;\mu_\pi^2|_{\rm toy} = 0.45\;  {\rm GeV}^2, \;\; \rho_{LS}^3|_{\rm toy} = 0.09\; {\rm GeV}^3, \;\;\rho_D^3|_{\rm toy} = 0.11  \; {\rm GeV}^3.
\end{equation}

Comparing with the values in Table~\ref{table:input}, we find $1-2\sigma$ shifts, with a rather poor fit quality. We note that this toy fit merely serves to illustrate how NP could be hidden in the HQE extraction, because the fit is rather flexible in accounting for such variations. Strong conclusions should not be made from this fit, except that it may be worth performing a full analysis on data. On the other hand, we also note that this may be challenging due to the large number of extra parameters.

\section{Forward-backward asymmetry}
\label{sec:AFB}
In this section, we consider the forward-backward asymmetry discussed in \cite{Turczyk:2016kjf} and more recently in \cite{Herren:2022spb}. 
The asymmetry is defined as
\begin{align}
    A_{FB} &\equiv \frac{\displaystyle\int_{-1}^0 \text{d}z \dfrac{\text{d} \Gamma}{\text{d}z}  - \int_{0}^1 \text{d}z \dfrac{\text{d} \Gamma}{\text{d}z} }{\displaystyle\int_{-1}^1 \text{d}z \dfrac{\text{d} \Gamma}{\text{d}z} } \ ,
    \label{eq::AFB}
\end{align}
where
\begin{align}
    z \equiv \cos\theta = \frac{v\cdot p_{\bar{\nu}_\ell} - v\cdot p_\ell}{\sqrt{(v\cdot q)^2 -q^2}} \ ,
\end{align}
and $\theta$ is the angle between spacial momenta of the lepton and the $B$ meson in the rest-frame of the dilepton pair. 

As discussed in~\cite{Herren:2022spb}, including a lepton energy cut $E_\ell^{\text{cut}}$ in the $A_{FB}$ definition leads to a cusp in the differential spectrum in the variable $z$, which can be 
problematic in experimental analysis.
To circumvent this issue, Ref.~\cite{Herren:2022spb} proposed to study 
$A_{FB}$ with a minimum cut on $q^2$ instead of $E_\ell$. We therefore consider only $q^2$ cuts, which also considerably simplifies the calculation. We refer to \cite{Herren:2022spb} for details of the calculation. 

Writing our results as in \eqref{eq::moments}, we find the $\xi$'s listed in Table~\ref{tab:forward-backward}. We consider for the first time the $\alpha_s$-corrections, both for the SM and for NP scenarios. 
In the upper part of Fig.~\ref{fig::dTdcos}, we show the differential distribution in $z$ normalized to $1/\Gamma_0$ as defined in Appendix~\ref{app:nptot} for the SM and our three NP scenarios in Table~\ref{table:NPScen}. Our normalization, i.e. using only $1/\Gamma_0$, differs from that used by  \cite{Turczyk:2016kjf,Herren:2022spb}, but our results for the SM are in agreement. 
In the lower panel of Fig.~\ref{fig::dTdcos}, we show the prediction for $A_{FB}$ as a function of $q^2_\mathrm{cut}$ where we plot the different SM contributions for illustration. The plots shows that forward-backward asymmetry and the differential distribution are sensitive the NP contributions and can distinguish among our three different scenarios.
The forward-backward asymmetry has not been measured so far, but our analysis shows the potential 
for understanding the SM and possibly to constrain NP contributions. 

\begin{table}[ht]
\begin{center}
\renewcommand{\arraystretch}{1.2} 
\begin{tabular}{c c }
\toprule
 & $A_{FB} \cdot 10^{-2}$   \\
\midrule
$\xi_{\text{SM}}$ & $24.603 |_\text{LO} -2.928 |_\text{pow}$ $- \left( \frac{\alpha_s}{\pi} \right) 6.63 $  \\
$\xi^{\braket{V_R ,V_R}}_{\text{NP}}$ & $-25.387|_\text{LO} + 0.769 |_\text{pow}  + \left( \frac{\alpha_s}{\pi} \right) 4.47 $  \\
$\xi^{\braket{S_L ,S_L}}_{\text{NP}}$ & $ -8.683|_\text{LO} -0.333 |_\text{pow} - \left( \frac{\alpha_s}{\pi} \right) 15.91 $  \\
$\xi^{\braket{S_R ,S_R}}_{\text{NP}}$ & $ -8.683|_\text{LO} -0.333 |_\text{pow} - \left( \frac{\alpha_s}{\pi} \right) 15.91 $  \\
$\xi^{\braket{T ,T}}_{\text{NP}}$ & $ -254.730|_\text{LO} +40.911 |_\text{pow} + \left( \frac{\alpha_s}{\pi} \right) 2.13 $ \\
$\xi^{\braket{V_L ,V_R}}_{\text{NP}}$ & $-24.208|_\text{LO} +4.025 |_\text{pow} +\left( \frac{\alpha_s}{\pi} \right) 7.53 $  \\
$\xi^{\braket{S_L ,S_R}}_{\text{NP}}$ & $ 12.104 |_\text{LO} - 1.415 |_\text{pow}- \left( \frac{\alpha_s}{\pi} \right) 24.67 $  \\
$\xi^{\braket{S_L ,T}}_{\text{NP}}$ & $49.207 |_\text{LO} +0.954|_\text{pow} + \left( \frac{\alpha_s}{\pi} \right) 51.17 $  \\
$\xi^{\braket{S_R , T}}_{\text{NP}}$ & $2.20 |_\text{pow}$  \\
\toprule
\end{tabular}
\caption{Numerical values of the parameters for the $A_{FB}$ given in Eq. (\ref{eq::moments}). We consider $q^2_{\text{cut}} = 4$ GeV$^2$. }
\label{tab:forward-backward}
\end{center}
\end{table}

\begin{figure}[htb]
	\centering
          \subfloat{\includegraphics[width=0.5 \textwidth]{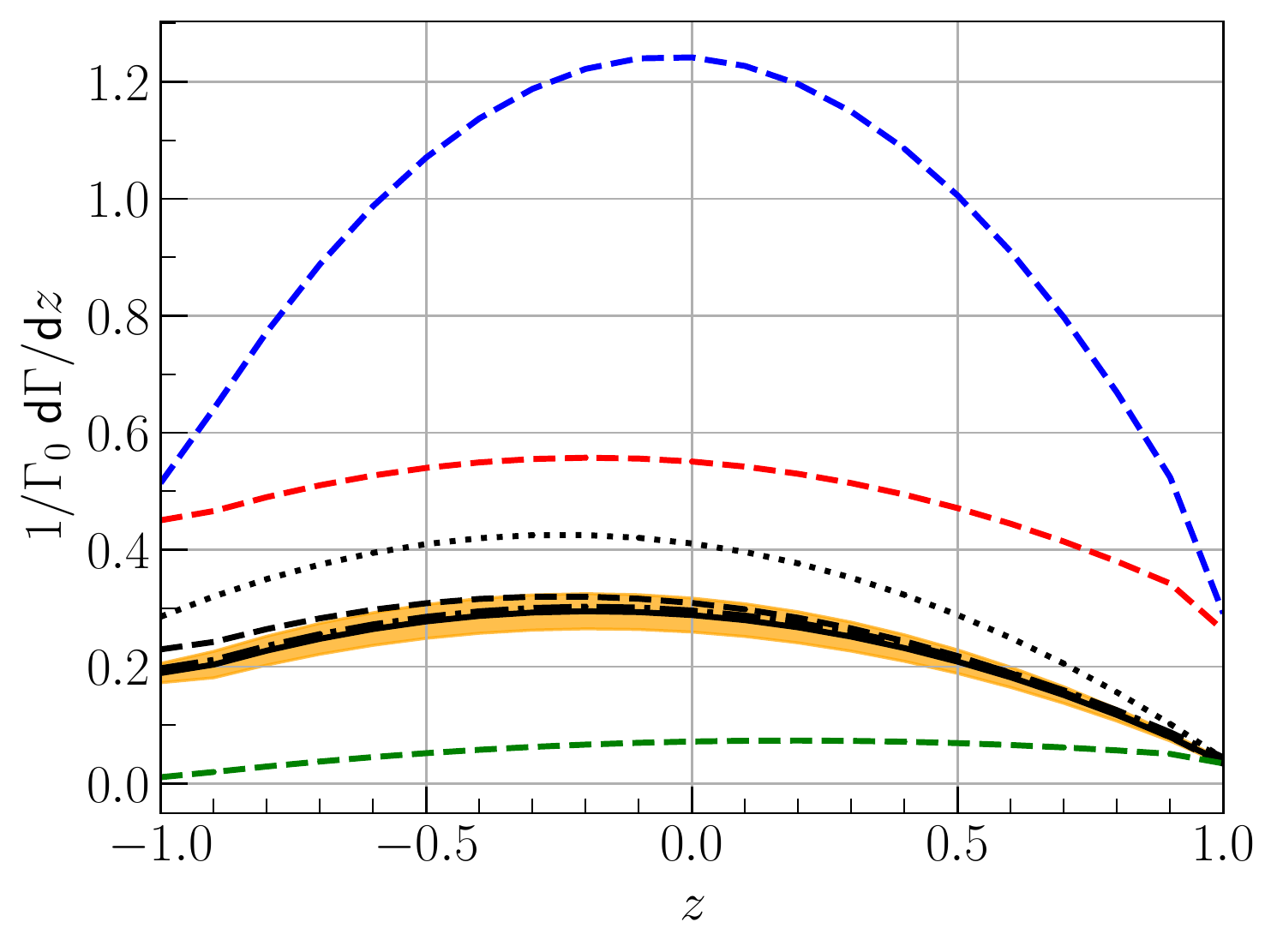}} 
          \subfloat{\includegraphics[width=0.5 \textwidth]{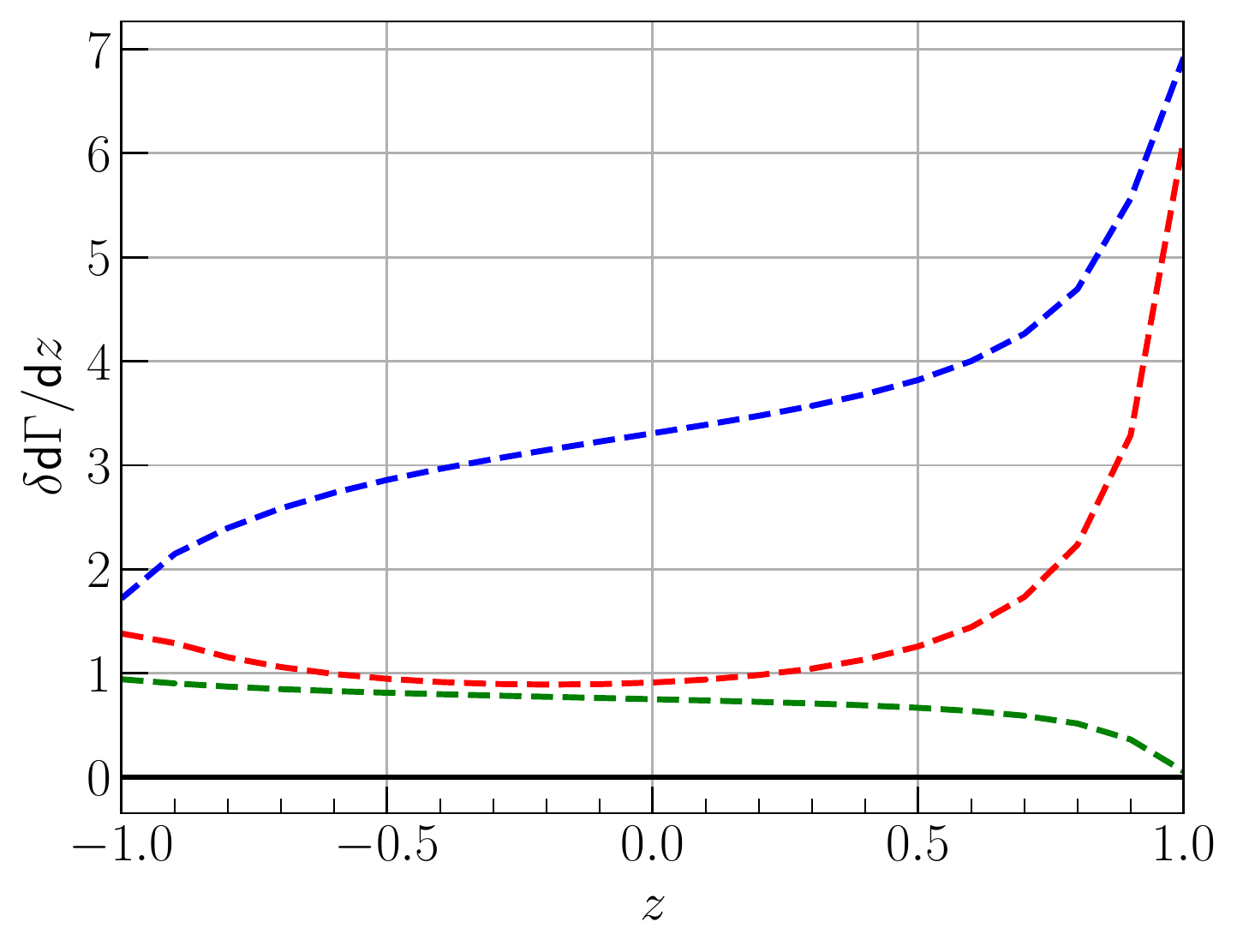}} \\
        \subfloat{\includegraphics[width=0.5 \textwidth]{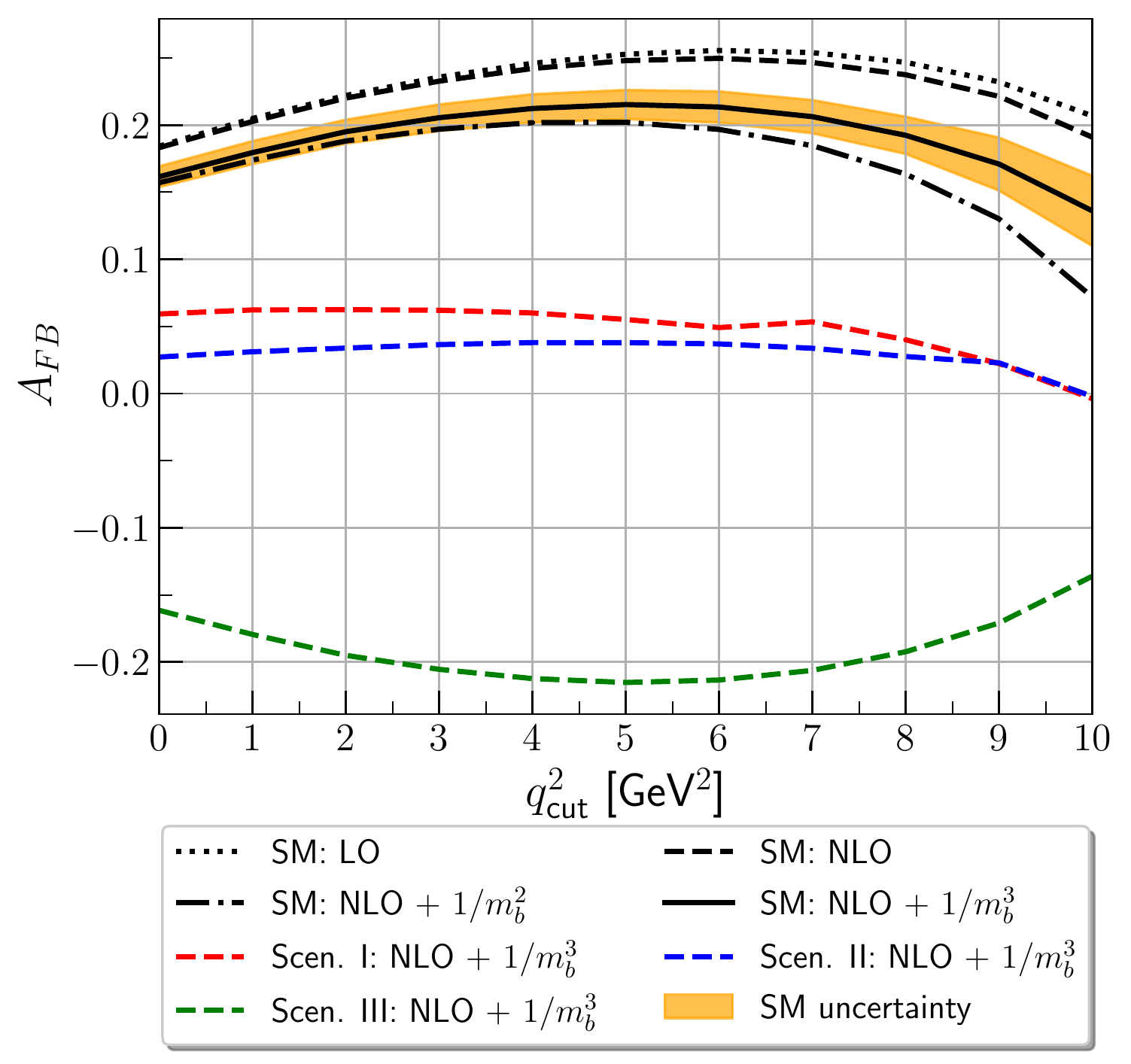}}
        \subfloat{\includegraphics[width=0.52 \textwidth]{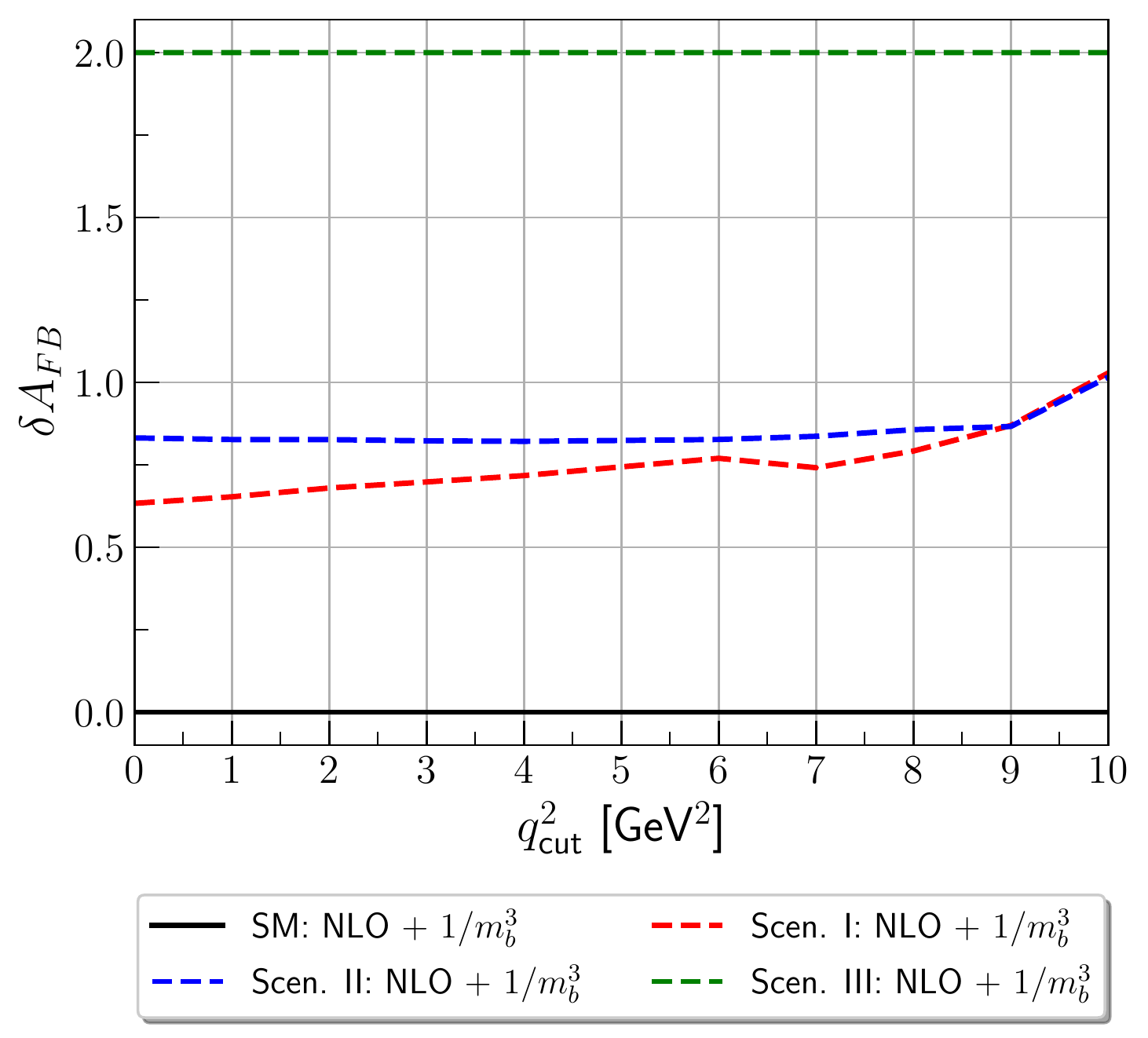}}
  \caption{(Upper part) The differential rate for $B \to X_c \ell \bar \nu_\ell$ as a function of $z$ without lepton energy cut 
  and relative size of the NP scenarios w.r.t.\ the SM prediction.
  (Lower part) Forward-backward asymmetry as a function of the $q^2$ cut for the three NP scenarios in Table~\ref{table:NPScen} and their relative size w.r.t.\ the SM.}
	\label{fig::dTdcos}
\end{figure}
\newpage

\section{Conclusion}
\label{sec::Conclusion}
We investigated New Physics effects on the semileptonic channel  $B \to X_c \ell \bar{\nu}_\ell$. For the first time, we compute power-corrections up to $\mathcal{O}(1/m_b^3)$ and $\alpha_s$-corrections for the full basis of the New Physics operators in the WET over the full differential decay width. These corrections are necessary to properly describe the dominant NP contributions to central moments of dilepton invariant mass $q^2$ and hadronic invariant mass $M_X^2$. 

We compared SM predictions, using HQE parameters obtained from experimental data, and experimental measurements to the moments of lepton energy, hadronic invariant mass and dilepton momentum for different toy New Physics scenarios. In addition, we also computed the forward-backward asymmetry. 
At the moment, it is challenging to compare the possible sensitivity to NP between the inclusive and exclusive channels, due to their different theoretical description. In the exclusive case, the hadronic form factors are known either from lattice or sum rules or a combination \cite{Bordone:2019guc, Bordone:2019vic,MILC:2015uhg,FermilabLattice:2021cdg}. As such, a measurement can in principle directly be translated into a constraint on NP (see \cite{Jung:2018lfu})\footnote{This also requires calculations for form factors that are not present in the SM}. Especially sensitive are angular analyses, which can be performed using normalized coefficients in which $|V_{cb}|$ drop out as in the inclusive case. A recent analysis on $B\to D^*$ data was done in \cite{Bordone:2019vic}, although there also some inconsistencies with the data were pointed out. For the inclusive case, the HQE parameter are currently fitted from experimental data. Therefore, a full NP analysis requires a simultaneous fit of the SM HQE parameters and NP coefficients. The main goal of this work is to pave the way for such a global analysis. Using all the available data to make optimal use of their complementary sensitivities to both the SM HQE parameters and the NP Wilson coefficients.

To further constrain such global fit, one may take advantage of lattice results for the HQE parameters, extracted from meson mass calculations at different quark mass values~\cite{Gambino:2017vkx}, and scattering matrix for $B\to X_c \ell \bar \nu_\ell$~\cite{Gambino:2020crt,Gambino:2022dvu}. Another approach could be to build upon the estimates for the HQE parameters in \cite{Hayashi:2021vdq}, where $\mu_G^2$ was obtained from the $B$-$B^*$ mass splitting and $\mu_\pi^2$ was obtained from a renormalon analysis \cite{Hayashi:2021vdq}. If the HQE parameters are known from theoretical calculations, even if preliminary, this could enhance the predictive power of the HQE by better assessing the non-perturbative inputs. In addition, the NP coefficients could then be directly determined from the data for which the $q^2$-moments seems most appropriate given their precision \cite{Belle:2021idw, Belle-II:2020oxx}. 

We aim to perform such a fit in a future publication using the EOS software \cite{EOSAuthors:2021xpv}. That analysis will then also determine whether the inclusive decays can compete with the exclusive channels in the search for NP.   

\subsubsection*{Acknowledgements}
We thank T.\ Mannel and M.\ Bordone for discussion and correspondence.
This  research  was supported by the Deutsche Forschungsgemeinschaft (DFG, German Research Foundation) under grant 396021762 - TRR 257.
The work of M.F.\ was supported by the European Union’s Horizon 2020 research and innovation program under the Marie Sk\l{}odowska-Curie grant agreement No.\ 101065445 -- PHOBIDE.

\pagebreak

\appendix
\section{NP contributions to the total rate}\label{app:nptot}
We decompose the prediction of the total rate in two parts:
\begin{align}
    \Gamma(B \to X_{c} \ell \bar{\nu}) &= \Gamma_0 \left( \Gamma^{\text{LO}}_{\text{NP}}(B \to X_{c} \ell \bar{\nu}) +\Gamma^{\text{Pow}}_{\text{NP}}(B \to X_{c} \ell \bar{\nu}) \right)
\end{align}
where
\begin{equation}
    \Gamma_0 = \dfrac{G_F^2 |V_{cb}|^2 m_b^5}{192 \pi^3} (1+ A_{\text{ew}})
\end{equation}
and $A_\text{ew}$ = 0.014~\cite{Sirlin:1974ni}.
The LO result in the free quark approximation is given by
\begin{align}
    \Gamma^{\text{LO}}_{\text{NP}}(B \to X_{c} \ell \bar{\nu}) &= \Gamma_{\text{SM}}^{\text{LO}}(B \to X_{c} \ell \bar{\nu}) \left(  |1 + C_{V_{L}}|^2 + |C_{V_{R}}|^2 + \frac{1}{4} \left( |C_{S_{L}}|^2 + |C_{S_{R}}|^2 \right) + 12 |C_{T}|^2 \right) \nonumber \\
    & + \Gamma^{\text{LO}}_{\text{mix}}(B \to X_{c} \ell \bar{\nu}) \left( \text{Re}( (1+ C_{V_{L}} ) C_{V_{R}} ) - \frac{1}{2} \text{Re} ( C_{S_{L}} C_{S_{R}} ) \right) \, ,
\end{align}
with
\begin{align}
    \Gamma^{\text{LO}}_{\text{SM}} &= (1 - 8 \rho -12 \rho^2 \log(\rho) + 8 \rho^3 - \rho^4) \, , \\
    \Gamma^{\text{LO}}_{\text{mix}} &= - 4 \sqrt{\rho} \, (1+ 9 \rho + 6\rho (1+\rho) \log(\rho) -9 \rho^2 -\rho^3 ) \ .
\end{align}
Our result agrees with the leading-order (LO) results from \cite{Jung:2018lfu}. 
The contribution from the power corrections is 
%cite{Colangelo:2016ymy} check result with our expression
\begin{align}
    \Gamma^{\text{Pow}}_{\text{NP}}&(B \to X_{c} \ell \bar{\nu}) = 
        %%=====================mupi====================%%
        \frac{\mu_\pi^2}{m_b^2} \Gamma^{\mu_\pi^2}_{\text{SM}} \left(|1+C_{V_L}|^2 + |C_{V_R}|^2 + \frac{1}{4} (|C_{S_L}|^2+|C_{S_R}|^2) + 12 |C_T|^2 \nonumber \right. \\ 
        & \left. - 2 \sqrt{\rho} \big( \rho^3 +9 \rho^2 -9 \rho -6(\rho+1) \rho \log(\rho) -1\big) \big( \text{Re}((1+C_{V_L}) C_{V_R}^*) \nonumber \right. \\
        & \left. - \frac{1}{2}  \text{Re}(C_{S_L} C_{S_R}^*) \big) \right)
        %%=====================mug-rhoLS====================%%
        + \left(\frac{\mu_G^2}{m_b^2}  - \frac{\rho_{LS}^3}{m_b^3} \right)  \left(\Gamma^{\mu_G^2}_{\text{SM}} \left( |1+C_{V_L}|^2 + |C_{V_R}|^2  \right) \nonumber \right. \\
        & \left. - \frac{1}{8} \big( 5 \rho^4 -32\rho^3 +72 \rho^2 -32 \rho  +12(\rho-4) \rho \log(\rho) -13 \big) (|C_{S_L}|^2 + |C_{S_R}|^2) \nonumber \right. \\
        & \left. -2 \big( 15 \rho^4 -64 \rho^3 +24 \rho^2  + 12(3\rho+4)\rho \log(\rho) +25 \big) |C_T|^2 \nonumber \right. \\
        & \left. + \frac{2 \sqrt{\rho}}{3} \big( 13 \rho^3-27\rho^2 -6 (3\rho^2-3\rho+2) \log(\rho)  +27 \rho -13 \big) \text{Re}( (1+C_{V_L}) C_{V_R}^*) \nonumber \right. \\
        & \left. - 3 \sqrt{\rho} \big( \rho^3-3\rho^2 -2(\rho^2 -5\rho -2) \log(\rho)  -9\rho +11 \big) \, \text{Re}( C_{S_L} C_{S_R}^* ) \right) \nonumber \\
        %%=====================rhoD====================%%
        & + \frac{\rho_D^3}{m_b^3} \left( \Gamma_{\text{SM}}^{\rho_D^3} \left( |1+C_{V_L}|^2 + |C_{V_R}|^2\right)  + \frac{1}{24} \big( -5 \rho^4 -8\rho^3 +12 (3\rho^2+8\rho+8) \log(\rho) \nonumber \right. \\ 
        & \left.  -184\rho +197 \big) \left( |C_{S_L}|^2 +|C_{S_R}|^2 \right) + 2 \big( -5 \rho^4 -8\rho^3 +32\rho^2 +4(9\rho^2-8\rho+8) \log(\rho) \nonumber \right. \\
        & \left. -56\rho +37 \big)  |C_T|^2 + 2 \left( \rho^3 -15\rho^2 +6(\rho^2 -\rho -2) \log(\rho) +39\rho -25 \right) \nonumber \right. \\
        & \left. \times \, \text{Re}((1+C_{V_L}) C_{V_R}^*) + \frac{2 \sqrt{\rho}}{6} \big( \rho^3 +9\rho^2 + (-18\rho^2 +90\rho +60) \log(\rho) \nonumber \right. \\
        & \left. -153\rho +143 \big) \, \text{Re}(C_{S_L} C_{S_R}^*) \right)
\end{align}
with
\begin{align}
    \Gamma^{\mu_\pi^2}_{\text{SM}} &= -\frac{1}{2}\Gamma_{\text{SM}}^{\text{LO}} \, , \\
     \Gamma^{\mu_G^2}_{\text{SM}} &= -\frac{1}{2} (5 \rho^4 -24\rho^3 +24\rho^2 +12\rho^2 \log(\rho) -8\rho +3) \, , \\
     \Gamma_{\text{SM}}^{\rho_{LS}^3} &= \frac{1}{2} \left( 5 \rho^4 -24 \rho^3 +24 \rho^2 +12 \rho^2 \log(\rho) -8\rho +3 \right) \, , \\
     \Gamma_{\text{SM}}^{\rho_D^3} &= \frac{1}{6} \left(-5 \rho^4 -8\rho^3 +24 \rho^2 +12(3\rho^2 +4) \log(\rho) -88\rho +77 \right)  \, .
\end{align}
For the power-corrections  $\mathcal{O}(1/m_b^2)$ of $(1+C_{V_L}) C_{V_R}$ our result agrees with \cite{Dassinger:2008as} $c_L c_R$ term.

\section{Repository}\label{sec:git}
The lengthy expressions for the various NP contributions to the inclusive moments are available in Mathematica format from the repository
\begin{verbatim}
    https://gitlab.com/vcb-inclusive/npinb2xclv
\end{verbatim}
We give predictions for the following quantities:
\begin{itemize}
    \item The first three moments of charged lepton energy with a lower cut on the charged lepton energy which are given by
    \begin{align}
        L_n = \frac{1}{\Gamma_0}
        \int_{E_\ell > E_\mathrm{cut}}
        \left( \frac{E_\ell}{m_b} \right)^n
        \frac{d \Gamma}{ d q^2 \, dE_\ell \, dE_\nu}
        \, dq^2 \, dE_\ell \, dE_\nu .
    \end{align}
    The expressions for $L_n$ are given in the files
    \verb|Ee_moment0.m, ..., Ee_moment3.m|.
    \item The first three moments of $M_X$ with a lower cut on the charged lepton energy which are given by
    \begin{align}
        M_n = \frac{1}{\Gamma_0}
        \int_{E_\ell > E_\mathrm{cut}}
        \left( \frac{M_B^2 + q^2 -2M_B q_0}{m_b^2} \right)^n
        \frac{d \Gamma}{ d q^2 \, dE_\ell \, dE_\nu}
        \, dq^2 \, dE_\ell \, dE_\nu .
    \end{align}
    The expressions for $M_n$ are given in the files \verb|MX_moment0.m, ..., MX_moment3.m|. 
    \item The first three moments of $q^2$ with a lower cut on the dilepton invariant mass which are given by
    \begin{align}
        Q_n = \frac{1}{\Gamma_0}
        \int_{q^2 > q^2_\mathrm{cut}}
        \left(\frac{q^2}{m_b^2} \right)^n
        \frac{d \Gamma}{ d q^2 \, dE_\ell \, dE_\nu}
        \, dq^2 \, dE_\ell \, dE_\nu .
    \end{align}
    The expressions for $Q_n$ are given in the files \verb|Q2_moment0.m, ..., Q2_moment3.m|. 
    \item The integrated total rate in the forward ($F$) and backward ($B$)directions with a lower cut on the dilepton invariant mass:
    \begin{align}
        \Gamma_F &= \frac{1}{\Gamma_0} 
     \int_0^1 dz \int_{q^2 > q^2_\mathrm{cut}} 
        \frac{d \Gamma}{ d q^2 \, du \, dz}, &
        \Gamma_B &= \frac{1}{\Gamma_0} 
    \int_{-1}^0   dz \int_{q^2 > q^2_\mathrm{cut}} 
        \frac{d \Gamma}{ d q^2 \, du \, dz}.     
    \end{align}
    The files \verb|Gamma_Forward.m| and \verb|Gamma_Backward.m| contain the expressions for $\Gamma_F$ and $\Gamma_B$.
    They can be then used to calculate the forward-backward asymmetry defined in Eq.~\eqref{eq::AFB}: $A_{FB} = (\Gamma_B-\Gamma_F)/(\Gamma_B+\Gamma_F)$. 
\end{itemize}
All analytic expressions are given using the on-shell scheme for the bottom and charm masses.
For the NP contributions, we include power corrections up to order $1/m_b^3$ and NLO corrections at the partonic level. The HQE parameters are denoted by the symbols \verb|mupi,muG,rhoD,rhoSL| while for the Wilson coefficients of the NP operators we use \verb|c[SL]|,\verb|c[SR]|,\verb|c[T]|,\verb|c[VL]|,\verb|c[VR]|. We use \verb|q2hatcut| to denote $q^2_\mathrm{cut}/m_b^2$ the lower cut on $q^2$ divided by the bottom mass squared. \verb|Ycut| corresponds to the ratio $2 E_\mathrm{cut} /m_b$.

The coefficient in front of $\alpha_s(\mu)/\pi$ is denoted by the functions \verb|X1El, X1mix, X1Q2| for $E_\ell, M_X$ and $q^2$ moments, respectively. The NLO correction to $\Gamma_F$ and $\Gamma_F$ is denoted by \verb|X1AFBForward| and \verb|X1AFBBackward|.
Since the evaluation of the NLO corrections involves Dirac delta functions and plus distributions, and a numerical integration over the phase-space, we provide  all necessary subroutines in the Mathematica package \verb|EvaluateAlphaSNP.m|. 
The package can be loaded in a Mathematica notebook with
\begin{verbatim}
In[] := << "alphas/EvaluateAlphaSNP.m"
\end{verbatim}
It defines the functions \verb|X1El, X1mix| and \verb|X1Q2| that execute the numerical phase-space integration of the $\alpha_s$ corrections. The NLO corrections the $Q_n$ moments are given by the function \verb|X1Q2[n,cNP,q2cuthat,m2,mu2hat]|, where:
\begin{itemize}
    \item \texttt{n}: the $n$th moment, with $0\le n \le 3$.
    \item \texttt{cNP}: product of NP coefficients appearing in the decay rate. Possible options are \verb|cNP = {SM^2,SM c[VL],SM c[VR],c[SL]c[SL],c[SR]c[SR],c[VL]^2,c[SL]c[SR],|
    \verb|c[VR]^2,c[VL]c[VR],c[SL]c[T],c[T]^2}|. \verb|SM^2| stands for the SM prediction.
    \item \verb|q2cuthat|: value of the lower cut in $q^2$ normalized to bottom mass squared.
    \item \verb|m2|: the mass ratio $m_c^2/m_b^2$.
    \item \verb|mu2hat|: value of the renormalization scale $\mu^2$ of the Wilson coefficients in the $\overline{\mathrm{MS}}$   scheme, normalized to the bottom mass squared.
\end{itemize}
The functions \verb|X1El| and \verb|X1mix| have similar syntax. 
For example, we find for the NLO corrections to the first $Q_1$ moment: 
\begin{verbatim}
In[]  := X1Q2[1, SM^2, 1/4.6^2, 1.15^2/4.5^2, 1]
Out[] := -0.215785
\end{verbatim}

\section{NP effects on the moments}
In this Appendix, we list the coefficients $\xi$ defined in \ref{eq::moments}. We categorize the contributions of leading-order, power-corrections and $\alpha_s$ corrections. 

\label{sec:appxi}
\begin{landscape}
\begin{table}[t]
\begin{center}
\renewcommand{\arraystretch}{1} 
\begin{tabular}{c c  c c}
\toprule
 & $\braket{E_\ell} \cdot 10^{-2} $ [GeV]&  $\braket{ (E_\ell - \braket{E_\ell})^2} \cdot 10^{-2}$ [GeV$^2$]&  $\braket{ (E_\ell - \braket{E_\ell})^3} \cdot 10^{-3}$ [GeV$^3$]  \\
\midrule
$\xi_{\text{SM}}$ & $157.23 |_\text{LO} -1.78 |_\text{pow}$ $-\left( \frac{\alpha_s}{\pi} \right) 4.608 $  & $8.715|_\text{LO} +0.291|_\text{pow}-\left( \frac{\alpha_s}{\pi} \right) 1.970 $ & $-3.076|_\text{LO}+3.399|_\text{pow}+\left( \frac{\alpha_s}{\pi} \right) 14.388 $ \\
$\xi^{\braket{V_R ,V_R}}_{\text{NP}}$ & $-10.00|_\text{LO} + 1.63|_\text{pow}  + \left( \frac{\alpha_s}{\pi} \right) 1.172 $  & $0.188|_\text{LO}-0.242|_\text{pow}+\left( \frac{\alpha_s}{\pi} \right) 0.531 $  & $9.394|_\text{LO}-1.555|_\text{pow} -\left( \frac{\alpha_s}{\pi} \right) 5.711 $ \\
$\xi^{\braket{S_L ,S_L}}_{\text{NP}}$ & $ 0.849|_\text{pow} + \left( \frac{\alpha_s}{\pi} \right) 0.414 $  & $0.128|_\text{pow}-\left( \frac{\alpha_s}{\pi} \right) 0.161  $  & $-0.712|_\text{pow}+\left( \frac{\alpha_s}{\pi} \right) 0.0365 $ \\
$\xi^{\braket{S_R ,S_R}}_{\text{NP}}$ & $0.849|_\text{pow} + \left( \frac{\alpha_s}{\pi} \right) 0.414 $ & $0.128|_\text{pow}-\left( \frac{\alpha_s}{\pi} \right) 0.161  $  & $-0.712|_\text{pow}+\left( \frac{\alpha_s}{\pi} \right) 0.0365 $ \\
$\xi^{\braket{T ,T}}_{\text{NP}}$ & $-77.960 |_\text{LO} +9.734|_\text{pow}  +\left( \frac{\alpha_s}{\pi} \right) 15.260 $  & $0.023 |_\text{LO} -1.120 |_\text{pow} +\left( \frac{\alpha_s}{\pi} \right) 6.887 $  & $71.831|_\text{LO} -7.041|_\text{pow}-\left( \frac{\alpha_s}{\pi} \right) 49.633 $ \\
$\xi^{\braket{V_L ,V_R}}_{\text{NP}}$ & $0.364|_\text{LO} -0.660|_\text{pow} +\left( \frac{\alpha_s}{\pi} \right) 2.864 $  & $-0.278|_\text{LO}-0.119|_\text{pow}-\left( \frac{\alpha_s}{\pi} \right) 0.462 $  & $-0.572|_\text{LO} -0.266|_\text{pow} -\left( \frac{\alpha_s}{\pi} \right) 0.672 $ \\
$\xi^{\braket{S_L ,S_R}}_{\text{NP}}$ & $ 0.182 |_\text{LO} +1.503|_\text{pow}+ \left( \frac{\alpha_s}{\pi} \right) 0.553 $ & $-0.139|_\text{LO} +0.234|_\text{pow} -\left( \frac{\alpha_s}{\pi} \right) 0.657 $ &  $-0.286|_\text{LO} -1.208|_\text{pow}+\left( \frac{\alpha_s}{\pi} \right) 0.315 $  \\
$\xi^{\braket{S_L ,T}}_{\text{NP}}$ & $9.745|_\text{LO}+0.575|_\text{pow}+\left( \frac{\alpha_s}{\pi} \right) 9.652 $ & $-0.0029|_\text{LO}+0.739|_\text{pow} -\left( \frac{\alpha_s}{\pi} \right) 0.279 $ &  $-8.978|_\text{LO}+ 1.011|_\text{pow} - \left( \frac{\alpha_s}{\pi} \right) 5.528 $ \\
$\xi^{\braket{S_R , T}}_{\text{NP}}$ & $0.624|_\text{pow}$  & $0.348|_\text{pow}$ & $0.780|_\text{pow}$ \\
\toprule
\end{tabular}
\caption{Numerical values of the coefficients $\xi$ in Eq.~\ref{eq::moments} for the lepton energy moments.
We consider $E_\ell^\mathrm{cut} = 1$ GeV.}
\label{tab:leptonmoments}
\end{center}
\end{table}

\begin{table}[H]
\begin{center}
\renewcommand{\arraystretch}{0.9} 
\begin{tabular}{c c  c c}
\toprule
 & $\braket{M_X} \cdot 10^{-1}$  [GeV$^2$]&  $\braket{ (M_X - \braket{M_X})^2} \cdot 10^{-1}$ [GeV$^4$]&  $\braket{ (M_X - \braket{M_X})^3} \cdot 10^{-1}$  [GeV$^6$]\\
\midrule
$\xi_{\text{SM}}$ & $43.016|_\text{LO} +0.0648|_\text{pow}+(\frac{\alpha_s}{\pi}) \, 7.219$ & $2.232|_\text{LO} +7.417|_\text{pow}+(\frac{\alpha_s}{\pi}) \, 29.666$  & $ -0.211|_\text{LO} +49.523|_\text{pow} -(\frac{\alpha_s}{\pi}) \, 53.141$\\
$\xi^{\braket{V_R , V_R}}_{\text{NP}}$ & $1.221|_\text{LO} -1.680|_\text{pow} -(\frac{\alpha_s}{\pi}) \, 0.925$ &$-0.554|_\text{LO}+3.466|_\text{pow} -(\frac{\alpha_s}{\pi}) \, 0.747$ & $0.019|_\text{LO} +1.586|_\text{pow} -(\frac{\alpha_s}{\pi}) \, 11.023$ \\
$\xi^{\braket{S_L , S_L}}_{\text{NP}}$ & $-0.600|_\text{LO} -0.7916|_\text{pow} -(\frac{\alpha_s}{\pi}) \, 2.041$ & $0.0421|_\text{LO}-0.853|_\text{pow} -(\frac{\alpha_s}{\pi}) \, 0.947$& $ 0.130|_\text{LO} -2.751|_\text{pow} -(\frac{\alpha_s}{\pi}) \, 5.180$\\
$\xi^{\braket{S_R , S_R}}_{\text{NP}}$ & $-0.600|_\text{LO} -0.7916|_\text{pow} -(\frac{\alpha_s}{\pi}) \, 2.041$  & $0.042|_\text{LO}-0.853|_\text{pow} -(\frac{\alpha_s}{\pi}) \, 0.947$ & $ 0.130|_\text{LO} -2.751|_\text{pow} -(\frac{\alpha_s}{\pi}) \, 5.180$\\
$\xi^{\braket{T , T}}_{\text{NP}}$ & $7.911|_\text{LO} +7.594|_\text{pow} +(\frac{\alpha_s}{\pi}) \, 1.926$ & $0.492|_\text{LO}+12.059|_\text{pow} +(\frac{\alpha_s}{\pi}) \, 10.068$ & $-1.620|_\text{LO} +39.365|_\text{pow} +(\frac{\alpha_s}{\pi}) \, 20.575$\\
$\xi^{\braket{V_L , V_R}}_{\text{NP}}$ & $-2.134|_\text{LO} +2.182|_\text{pow} -(\frac{\alpha_s}{\pi}) \, 0.274$ & $0.364|_\text{LO}-5.449|_\text{pow} -(\frac{\alpha_s}{\pi}) \, 0.137$ & $0.404|_\text{LO} -8.685|_\text{pow} +(\frac{\alpha_s}{\pi}) \, 5.821$\\
$\xi^{\braket{S_L , S_R}}_{\text{NP}}$ & $-1.067|_\text{LO}-1.616|_\text{pow} -(\frac{\alpha_s}{\pi}) \, 3.189$ & $0.182|_\text{LO} -0.811|_\text{pow}-(\frac{\alpha_s}{\pi}) \, 1.786$ & $0.202|_\text{LO} -4.346|_\text{pow} -(\frac{\alpha_s}{\pi}) \, 9.340$\\
$\xi^{\braket{S_L , T}}_{\text{NP}}$ & $0.213|_\text{LO}-0.215|_\text{pow} +(\frac{\alpha_s}{\pi}) \, 0.890$ & $-0.145|_\text{LO}+1.040|_\text{pow} +(\frac{\alpha_s}{\pi}) \, 0.018$ &  $-0.058|_\text{LO} +0.656|_\text{pow} +(\frac{\alpha_s}{\pi}) \, 3.876$\\
$\xi^{\braket{S_R , T}}_{\text{NP}}$ & $-0.081|_\text{pow}$ & $0.327|_\text{pow}$ & $0.193|_\text{pow}$\\
\toprule
\end{tabular}
\caption{Numerical values of the coefficients $\xi$ in~\ref{eq::moments} for the hadronic invariant mass moments. 
We consider $E_\ell^\mathrm{cut} = 1$~GeV.}
\label{tab:Mxmoments}
\end{center}
\end{table}

\begin{table}[t]
\begin{center}
\renewcommand{\arraystretch}{1.2} 
\begin{tabular}{c c  c c}
\toprule
 & $\braket{q^2} $ [GeV$^2$]&  $\braket{ (q^2 - \braket{q^2})^2} $ [GeV$^4$]&  $\braket{ (q^2 - \braket{q^2})^3} $ [GeV$^6$] \\
\midrule
$\xi_{\text{SM}}$ & $7.072|_\text{LO} -0.449|_\text{pow}+(\frac{\alpha_s}{\pi}) \, 0.168$ &  $4.278|_\text{LO} -1.727 |_\text{pow} + (\frac{\alpha_s}{\pi}) \, 0.854$  & $3.773|_\text{LO}-4.695|_\text{pow} +(\frac{\alpha_s}{\pi}) \, 6.879$\\
$\xi^{\braket{V_R , V_R}}_{\text{NP}}$ & $-0.681|_\text{LO}+0.138|_\text{pow} +(\frac{\alpha_s}{\pi}) \, 0.121$ & $-1.231|_\text{LO} +0.429|_\text{pow} +(\frac{\alpha_s}{\pi}) \, 1.467$ & $0.486|_\text{LO} -0.136|_\text{pow} +(\frac{\alpha_s}{\pi}) \, 5.182$ \\
$\xi^{\braket{S_L , S_L}}_{\text{NP}}$ & $0.135|_\text{LO}+0.182|_\text{pow} + (\frac{\alpha_s}{\pi}) \, 0.373$ & $0.099|_\text{LO} +0.592|_\text{pow} +(\frac{\alpha_s}{\pi}) \, 0.379$& $-0.512|_\text{LO} +1.236|_\text{pow} -(\frac{\alpha_s}{\pi}) \, 0.789$\\
$\xi^{\braket{S_R , S_R}}_{\text{NP}}$ & $0.135|_\text{LO}+0.182|_\text{pow} + (\frac{\alpha_s}{\pi}) \, 0.373$ & $0.099|_\text{LO} +0.592|_\text{pow} +(\frac{\alpha_s}{\pi}) \, 0.379$& $-0.512|_\text{LO} +1.236|_\text{pow} -(\frac{\alpha_s}{\pi}) \, 0.789$\\
$\xi^{\braket{T , T}}_{\text{NP}}$ & $-2.174|_\text{LO}+0.510|_\text{pow} +(\frac{\alpha_s}{\pi}) \, 0.535$ & $-1.591|_\text{LO} +1.290|_\text{pow} -(\frac{\alpha_s}{\pi}) \, 0.275$ & $8.200|_\text{LO}-1.716|_\text{pow} -(\frac{\alpha_s}{\pi}) \, 10.168$\\
$\xi^{\braket{V_L , V_R}}_{\text{NP}}$ & $0.692|_\text{LO}-0.108|_\text{pow} -(\frac{\alpha_s}{\pi}) \, 0.090$ & $0.765|_\text{LO} -0.248|_\text{pow} -(\frac{\alpha_s}{\pi}) \, 1.327$ & $-2.109|_\text{LO}+0.814|_\text{pow} -(\frac{\alpha_s}{\pi}) \, 2.351$\\
$\xi^{\braket{S_L , S_R}}_{\text{NP}}$ & $0.346 |_\text{LO} +0.359|_\text{pow} +(\frac{\alpha_s}{\pi}) \, 0.759$ & $0.382|_\text{LO} +1.152|_\text{pow} +(\frac{\alpha_s}{\pi}) \, 0.557$ & $-1.05|_\text{LO} +2.492|_\text{pow} -(\frac{\alpha_s}{\pi}) \, 2.439$\\
$\xi^{\braket{S_L , T}}_{\text{NP}}$ & 0 & 0 &  0 \\
$\xi^{\braket{S_R , T}}_{\text{NP}}$ & $0$ & $0$ & $0$\\
\toprule
\end{tabular}
\caption{Numerical values of the coefficients $\xi$ in Eq.~\ref{eq::moments} for the $q^2$ moments. 
We consider $q^2_\mathrm{cut} = 4$ GeV$^2$.}
\label{tab:q2moments}
\end{center}
\end{table}
\end{landscape}

\bibliographystyle{jhep} 
\bibliography{main.bib} 

\end{document}